\documentclass[aps,prb,twocolumn,amsmath,amssymb,superscriptaddress,showpacs]{revtex4-1}

\usepackage[utf8]{inputenc}
\usepackage{amsfonts}
\usepackage{amsmath}
\usepackage{amssymb}
\usepackage{amsthm}
\usepackage{fontenc}
\usepackage{graphicx}
\usepackage{xcolor}
\usepackage{textcomp}
\usepackage{epstopdf}
\usepackage{braket}
\usepackage{mathtools}
\usepackage{amsmath}
\usepackage{dcolumn}
\usepackage{multirow}

\begin{document}
\newcommand{\abs}[1]{\lvert#1\rvert}
\title{ Symmetries of quantum transport with Rashba spin-orbit: Graphene spintronics}

\author{Leonor Chico}
\affiliation{Instituto de Ciencia de Materiales de Madrid, Consejo Superior de Investigaciones Cient\'{\i}ficas (ICMM-CSIC),  C/ Sor Juana In\'es de la Cruz 3, 28049 Madrid, Spain}
\author{A. Latg\' e}
\affiliation{Instituto de F\' isica, Universidade Federal Fluminense, Niter\' oi, Av.~Litor\^anea sn 24210-340, RJ-Brazil}
\author{Luis Brey}
\affiliation{Instituto de Ciencia de Materiales de Madrid, Consejo Superior de Investigaciones Cient\'{\i}ficas (ICMM-CSIC), C/ Sor Juana In\'es de la Cruz 3, 28049 Madrid, Spain}

\date{\today}

\begin{abstract}
The lack of some spatial symmetries in planar devices with Rashba spin-orbit interaction opens the possibility of producing spin polarized electrical currents in absence of external magnetic field or magnetic impurities. We study  how the direction of the spin polarization of the current is related to spatial symmetries of the device. As an example of these relations we study numerically the spin-resolved current in graphene nanoribbons. Different configurations are explored and analyzed to demonstrate that graphene nanoflakes may be used as excellent spintronic devices in an all-electrical setup.

\end{abstract}

\maketitle

\section{Introduction}

One of the most important challenges in physics and materials science is the exploration of novel systems and physical mechanisms for 
spintronics, with the aim of designing high-speed and low-power devices.\cite{Wolf2001,Zutic2004,Han2014,Tang2015} 
 In particular, the production and detection of spin-polarized currents by electrical means is a newly explored route towards this goal. 
 Remarkably, spin-dependent transport can be achieved in systems with spin-orbit interaction (SOI) in the absence of ferromagnetic contacts or external magnetic fields.\cite{Debray2009} 
The extrinsic Rashba SOI couples the orbital motion and the spin of the electron under an external electric field, allowing for the manipulation of spins without breaking time reversal symmetry.\cite{WinklerBook} 
The possibility to tune SOI in Rashba systems is increasing the exploration of materials and devices which exploit this effect. 

Devices and materials with Rashba SOI are intensely studied for spintronic applications, such as transition metal dichalcogenides, (TMC) \cite{Yuan2013,Yuan2014} surfaces or novel two-dimensional (2D) nanosheets,\cite{Yin2013,Tsai2013,Ma2014} and one-dimensional systems, i.e., nanowires.\cite{Liang2012,Reuther2013} 
Theoretically, spin transport has been studied in semiconductor quantum wires with Rashba SOI and magnetic field modulations,\cite{Feng2005, Zhang2007, Xu2013, Jelena2013} and there are theoretical proposals for spin filters based in 2D graphene\cite{David2014} and graphene nanoribbons with specific geometries.\cite{Liu2012,Zhang2013} 
 Indeed, the experimental ability to fabricate precise graphene nanoribbons (GNRs) by bottom-up fabrication processes\cite{Huang2012,Yuan2013} or by epitaxial growth on silicon carbide\cite{Baringhaus2014} 
signals these ribbons as potentially fundamental building blocks in nanoelectronics and is undoubtedly one of the reasons to suggest them as spin-orbit-based devices.

In pristine graphene the intrinsic SOI is negligible;\cite{Huertas2007,Min2007} however, other spin-orbit couplings  induced by different mechanisms 
such as hydrogenation, 
chemical functionalization or
proximity effect with materials with strong SOI,
have been theoretically proposed \cite{CastroNeto2009,Weeks2011,Gmitra2013,Eremeev2014} and experimentally realized.\cite{Ma2012,Marchenko2012,Balakrishnan2013,Avsar2014,Calleja2015} 
The experimentally reported enhancement of SOI in graphene due to weak hydrogenation,\cite{Balakrishnan2013} gold hybridization,\cite{Marchenko2012} or proximity to WS$_2$ \cite{Avsar2014} is of three orders of magnitude or larger, 
indicating the possible use of graphene in spintronic devices. Recently, a giant spin-Hall effect has been experimentally measured in graphene, 
due to the dramatic increase of SOI produced by Cu atoms on graphene grown by chemical vapor deposition, reporting SOI splittings around 20 meV.\cite{Balakrishnan2014} Intercalation of Au atoms in graphene grown on Ni has lead to SOI splittings around 100 meV due to hybridization with gold atoms. \cite{Marchenko2012} Calleja {\it et al.} report larger values when Pb is intercalated between graphene and the Ir substrate.\cite{Calleja2015} In addition, spin angle-resolved photoemission spectroscopy experiments suggest that a large Rashba-type SOI can be  
tuned in graphene by the application of an external electric field:\cite{Varykhalov2008,Dedkov2008} in samples grown on Ni(111), splittings larger than 200 meV have been reported.\cite{Dedkov2008} 
In nanotubes and curved graphene, hybridization between $\pi$ and $\sigma$ bands induces a SOI effect larger than in graphene,\cite{Chico2004} so folds and wrinkles may also increase SOI in graphene systems.\cite{Costa2013}  Therefore, altering graphene by hybridization is an active route to achieve SOI values of interest for spintronics.

{\it In this work we show that the existence or absence of spin-polarized currents
in planar devices can be predicted based on a combination of time reversal and spatial symmetries.} 
We give a global and comprehensive vision of the possible spin polarization directions for planar geometries in Rashba systems. 
As an interesting example, we discuss here the spin-dependent conductance of different GNRs, which also present electron-hole symmetry. 
We show that anti-zigzag and anti-armchair GNRs with an electric-field-induced Rashba coupling in the central region,
can produce a spin-polarized current in the direction perpendicular to the ribbon, without the need of breaking time reversal symmetry. Furthermore, we also demonstrate that spin-polarized currents in the transversal direction can appear in any nanoribbon 
of constant width. 

Indisputably, the discussed effects will be larger in other materials, such as TMC \cite{Yuan2013} or germanene and stannene nanoribbons,\cite{Yin2013} but we choose graphene as a prototype system to study SOI effects, which can be modeled with a simple Hamiltonian,\cite{Zhang2014} with the idea that symmetry reasoning is equally applicable to other materials which require a more sophisticated calculation. Notwithstanding, given the giant SOI experimentally reported in graphene,\cite{Ma2012,Marchenko2012,Balakrishnan2013,Balakrishnan2014,Avsar2014,Calleja2015} 
we propose the use of {\it perfect graphene nanoribbons as magnetic-field-free spintronic devices}.

\section{Geometry and definitions}

\begin{figure}[hbt]
\includegraphics[width=0.94\columnwidth]{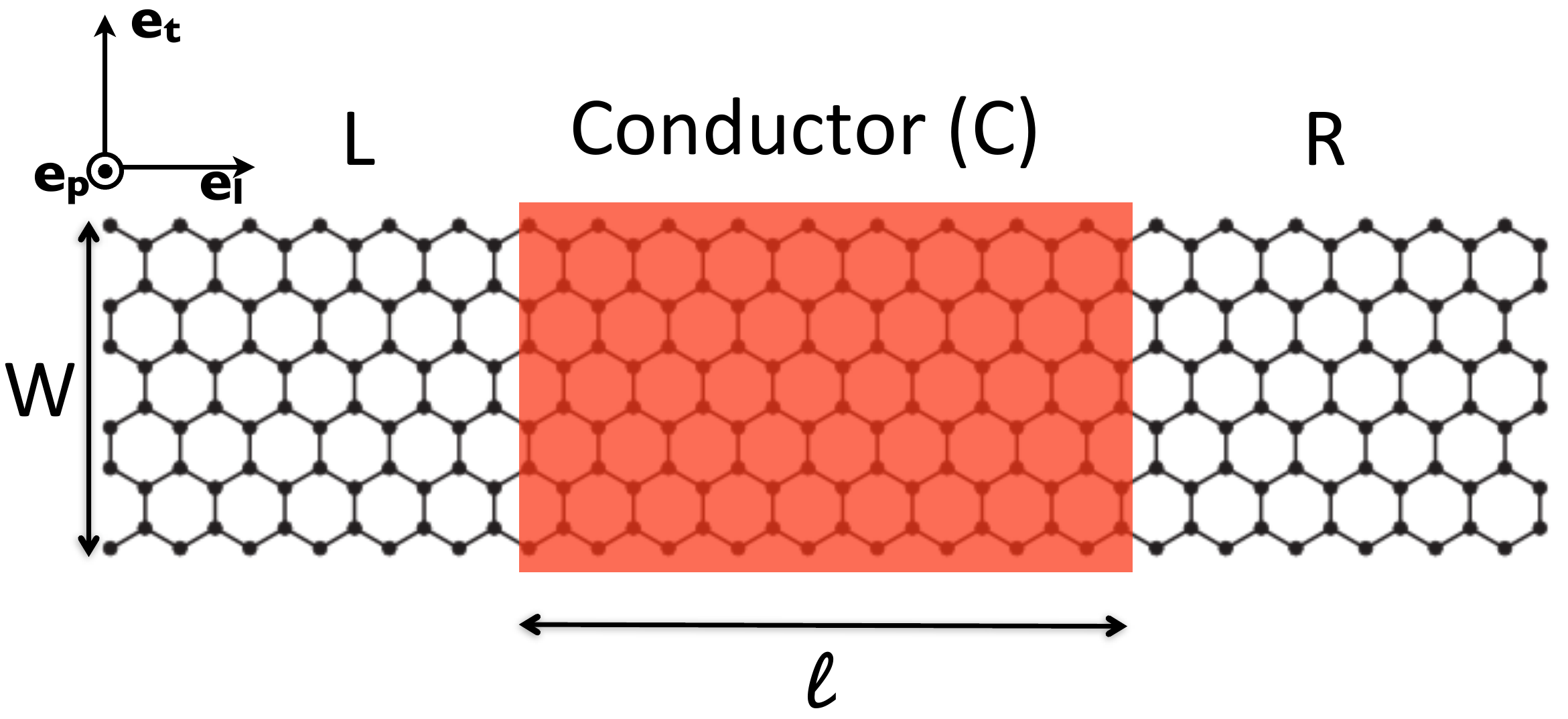}\caption{(Color online) Schematic drawing of the device geometry.
Left (L) and right (R) contacts 
are pristine nanoribbons without SO interaction. The conductor (C), shaded in red, is the central part of the device with Rashba SOI, length $\ell$, and width $W$.}
\label{device}
\end{figure}

We consider planar two-terminal devices as the one shown in Fig. \ref{device}. The current flows between left ($L$) and right ($R$) contacts along the longitudinal $\bm{e_l}$-direction. The spin-orbit coupling occurs in the
central part of the device  that is
perturbed by a Rashba-like SOI generated by an electric field applied perpendicular to the ribbon; i.e., in the $\bm{e_p}$-direction.  The unitary vector $\bm{e_t}$ defines the transverse direction of the ribbon, i.e., across its width. The conductance $\textit{G}^{LR}_{\sigma \sigma'} (E)$ indicates the probability that an electron in the left electrode with energy $E$ and spin pointing in the $\sigma$-direction reaches the right electrode with spin pointing in the 
 $\sigma'$-direction.

\section{Symmetry considerations}

In planar quasi-one-dimensional devices as those considered here, one can expect the following spatial symmetries: 
\par \noindent 
(i) $C_2$ rotation around $\bm{e_p}$. Under $C_2$, the spatial, momentum and spin components change as 
$(e_l,e_t,e_p\rightarrow -e_l,-e_t,e_p)$, $(p_l,p_t,p_p\rightarrow -p_l,-p_t,p_p)$, and $(\sigma_l,\sigma_t,\sigma_p \rightarrow -\sigma_l,-\sigma_t,\sigma_p )$. Therefore, 
the conductance of the device is invariant under these operations. This amounts to interchange the left and right electrodes, and invert the spin direction along the $\bm{e_l}$ or $\bm{e_t}$ directions, i.e., $G^{LR}_{\sigma \sigma'}= G^{RL}_{\bar{\sigma}\bar{\sigma}'}$,  where $\bar{\sigma}$ indicates a spin projection opposite to $\sigma$, and $\sigma'$=$\pm \sigma$.
For the spin direction perpendicular to the device, we get $G^{LR}_{\sigma \sigma'}= G^{RL}_{\sigma \sigma'}$. 
\par \noindent 
(ii) Longitudinal mirror symmetry $M_l$. 
For $M_l$, the spatial and momentum components transform as $(e_l,e_t,e_p\rightarrow e_l,-e_t,e_p)$ and $(p_l,p_t,p_p\rightarrow p_l,-p_t,p_p)$, respectively. For the spin components, recalling that the spin transforms as an axial vector, we have $(\sigma_l,\sigma_t,\sigma_p \rightarrow -\sigma_l,\sigma_t,-\sigma_p )$. Thus, this symmetry does not change the roles of the electrodes, but changes the sign of the spin projection when the spin direction is either 
$\bm{e_l}$ or $\bm{e_p}$, 
leading to the relation $G^{LR}_{\sigma \sigma'}= G^{LR}_{\bar{\sigma}\bar{\sigma}'}$. Notice that this symmetry does not give any relationship for the conductance when the spin direction is along $\bm{e_t}$.
\par \noindent
(iii) Transversal mirror symmetry $M_t$.
Under $M_t$, the spatial and momentum components transform as $(e_l,e_t,e_p\rightarrow -e_l,e_t,e_p)$, $(p_l,p_t,p_p\rightarrow -p_l,p_t,p_p)$, whereas the spin 
changes as $(\sigma_l,\sigma_t,\sigma_p \rightarrow \sigma_l,-\sigma_t,-\sigma_p )$. Therefore, the relation $G^{LR}_{\sigma \sigma'}= G^{RL}_{\bar{\sigma}\bar{\sigma}'}$ is obtained when the spin is pointing in the $\bm{e_t}$ or $\bm{e_p}$ directions. Otherwise, for the spin pointing in the longitudinal direction $\bm{e_l}$, we obtain the relation $G^{LR}_{\sigma \sigma'}= G^{RL}_{\sigma \sigma'}$. 

\begin{table}[ht]
\renewcommand{\arraystretch}{2}
\setlength{\tabcolsep}{6pt} 
\caption{Symmetries and the corresponding conductance relations derived from them. First column: symmetries; second column: spin projection directions; third column: spin-resolved conductance relations.  In this Table $\sigma'$=$\pm \sigma$. }
\begin{center}
\begin{tabular}{|c|c|c|}
   \hline
      $\Theta$ & $ \bm{e_l}, \bm{e_t}, \bm{e_p}$ & $G^{LR}_{\sigma \sigma'} = G^{RL}_{\bar{\sigma}' \bar{\sigma}}$ \\
   \hline
      $U$ & $ \bm{e_l}, \bm{e_t}, \bm{e_p}$ & $G^{LR}_{\sigma \sigma'} (E)= G^{RL}_{{\sigma}' {\sigma}}(-E)$ \\
      \hline
         $M_l$ & $ \bm{e_l}, \bm{e_p}$ & $G^{LR}_{\sigma \sigma'} = G^{LR}_{\bar{\sigma} \bar{\sigma}'}$ \\
       \hline
     \multirow{2}{*}{$M_t$}& $\bm{e_l}$  &  $G^{LR}_{\sigma \sigma'} = G^{RL}_{{\sigma} {\sigma}'}$  \\
     \cline{2-3}
                    & $\bm{e_t}, \bm{e_p}$ & $G^{LR}_{\sigma \sigma'} = G^{RL}_{\bar{\sigma} \bar{\sigma}'}$\\
    \hline
     \multirow{2}{*}{$C_2$}& $\bm{e_l}, \bm{e_t}$  &  $G^{LR}_{\sigma \sigma'} = G^{RL}_{\bar{\sigma} \bar{\sigma}'}$  \\
     \cline{2-3}
                    & $\bm{e_p}$ & $G^{LR}_{\sigma \sigma'} = G^{RL}_{{\sigma} {\sigma}'}$\\
\hline
\end{tabular}
\end{center}
\label{TS}
\end{table}

Besides spatial symmetries, in the absence of magnetic field time reversal symmetry ($\Theta$) holds, implying  
$G^{LR}_{\sigma \sigma'}= G^{RL}_{\bar{\sigma}'\bar{\sigma}}$.  
Finally, in presence of  electron-hole symmetry ($U$), the conductance as a function of the energy $E$ satisfies $G^{LR}_{\sigma \sigma'}(E)= G^{RL}_{{\sigma}'{\sigma}}(-E)$. All these relations are gathered in  Table \ref{TS}. 
These symmetry rules  allow us to predict  the possibility of obtaining spin-polarized currents in any planar devices.
For an incident unpolarized current from the left electrode,
the spin polarization of the current in the right electrode in the $s$-direction $(s=l,t,p)$ is defined as
\begin{equation}
P_s = G^{LR}_{ss} + G^{LR}_{\bar s s} -G^{LR}_{\bar s \bar s} - G^{LR}_{s \bar s } \, \, .
\label{Pola}
\end{equation}
Therefore, from symmetry considerations we obtain that in systems with $M_l$ symmetry the spin polarization of the current in the $\bm{e_l}$ and $\bm{e_p}$ spin-directions are zero, whereas in all the other cases a spin polarized current can be obtained, albeit with different intensities and for various reasons.

\begin{figure}[h]
\includegraphics[width=\columnwidth]{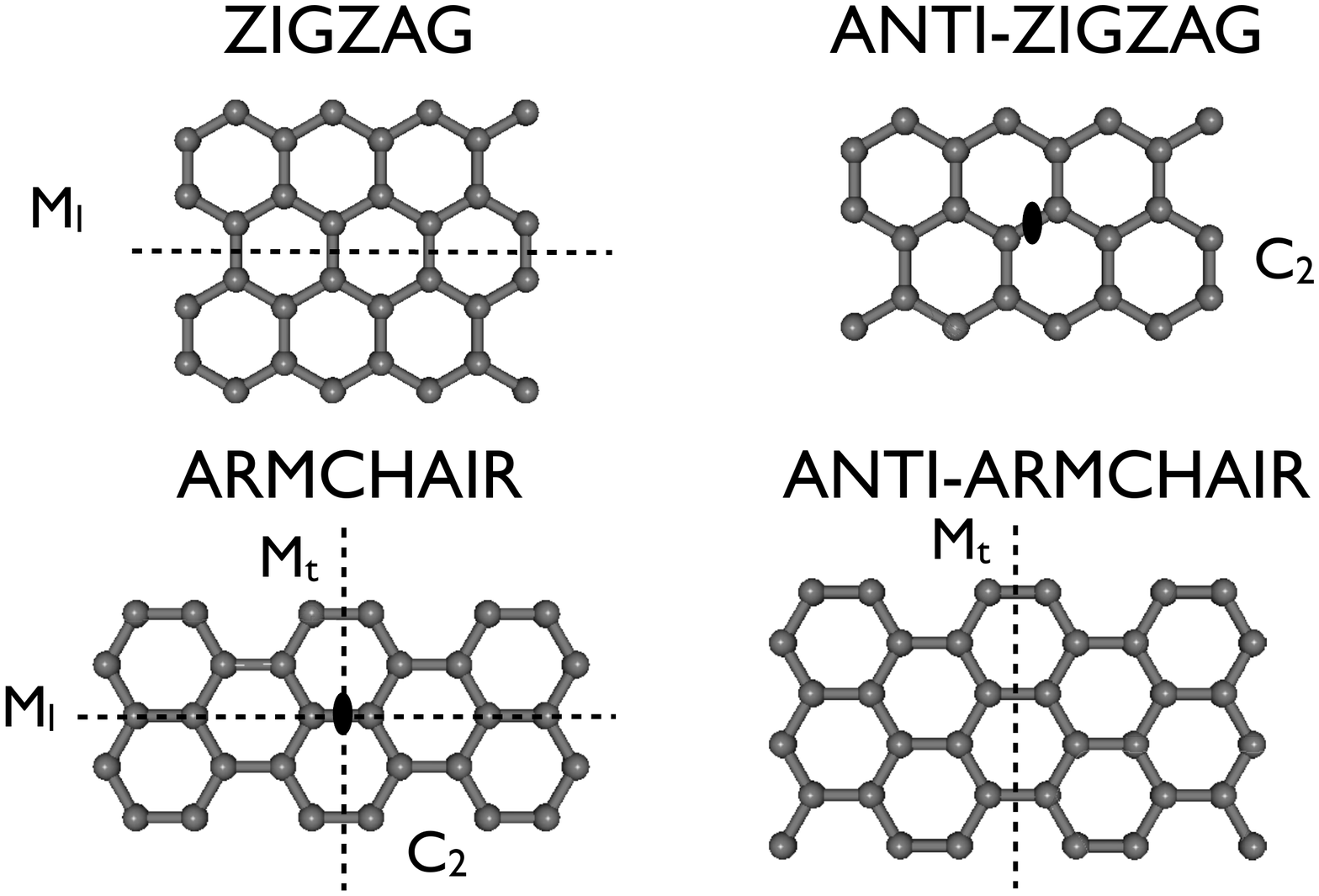}
\caption{Schematic drawing of the four considered central conductors and their relevant spatial symmetries.
\label{figflakes}
}
\end{figure}

\section{Spin-polarized current in graphene nanoribbons}

We consider the simplest possible geometry: a graphene nanoribbon where the central region has a finite Rashba SOI.
The length of the conductor $\ell$ is given by $3 a_{cc}N$ for an armchair (AC) GNR and by $\sqrt{3}a_{cc}N$ for a zigzag (ZZ) one, with $N$ being the number of longitudinal unit cells and $a_{cc}$ the carbon bond length in graphene.  
The ribbon width, $W$, is defined by the number of dimers (zigzag lines) across the width of the armchair (zigzag) nanoribbon, given hereafter by $M$.  As the focus is on transport properties, in the case of AC terminations we restrict the study to metallic ribbons, i.e., those with $M=3q+2$, $q$ being an integer.\cite{Brey2006}

 \begin{figure}[h]
\includegraphics[width=\columnwidth]{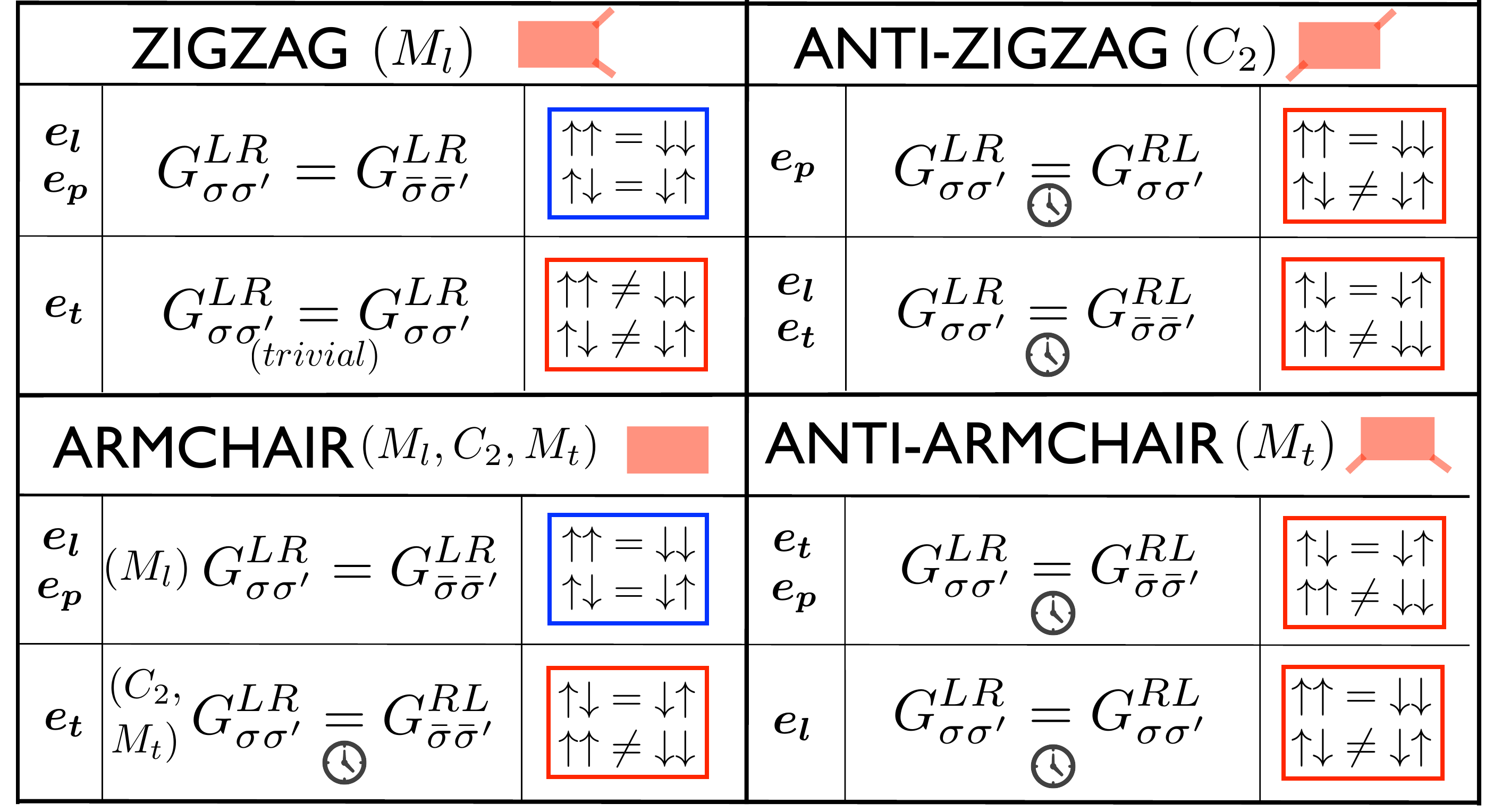}\caption{Graphical summary of our results for graphene nanoribbon flakes. For each case, the first column lists the spin direction; the second column shows the corresponding  spin-resolved conductance relation with the spatial symmetry employed for its derivation, indicating also whether time-reversal symmetry is needed (with a clock icon) in order to obtain the final result shown in the third column, i.e., which spin conductances are equal or not. Up and down arrows are referred to the spin projection direction indicated in the first column.
In the last column the red (blue) color of the rectangle surrounding the final spin conductance relations indicates the possibility (impossibility) of getting spin-polarized transport.}
\label{conductable}
\end{figure}

Infinite zigzag GNRs have transversal mirror symmetry $M_t$ for any $M$, but this symmetry is not present in a rectangularly cut finite-size flake of length $N$, such as those depicted in the upper panels of  Fig.\ref{figflakes}.  The parity of $M$ defines two kinds of zigzag flakes:  even-$M$ zigzag GNRs that have a longitudinal mirror symmetry $M_l$ and  odd-$M$ zigzag GNRs which are invariant under $C_2$.
The $M$-odd zigzag ribbons are commonly called anti-zigzag (AZZ), as shown in Fig. \ref{figflakes}.   
 For armchair GNRs, the more symmetric configurations, with $M_l$, $M_t$ and $C_2$ symmetries, happen for $M$ odd. $M$-even AC GNRs only have $M_t$, both in the infinite case and in the finite flake and they are called anti-armchair (AAC) ribbons, see Fig. \ref{figflakes}.  
 
The application of the symmetry relations summarized in Table \ref{TS} to the GNR flakes depicted in Fig. \ref{figflakes} tells us the possibility of obtaining spin-polarized currents in graphene-based devices. 
Fig. \ref{conductable} shows the relation between the spin-resolved conductances for different graphene flakes.  
Note that for $M_t$ and $C_2$ symmetries, it is necessary to consider also time reversal symmetry $\Theta$ to obtain the relations between the spin-polarized currents flowing in the same direction, as indicated in Fig. \ref{conductable}. 
The same symmetry arguments can be easily generalized along 
the lines described in this work to other graphene flake geometries. However, the use of the symmetry relations does not give us information on the magnitude of the spin polarization. In order to quantify the spin polarization it is necessary to perform microscopic calculations taking into account all the details of the discrete lattice and distinguishing between different symmetries. To do that we use a tight-binding formalism for computing the conductance of different GNRs.

\subsection{Microscopic calculations}

The Rashba spin-orbit interaction in graphene 
can be described in the nearest-neighbor hopping tight-binding approximation \cite{Qiao2010,Lenz2013}. The total Hamiltonian is $H=H_0 + H_R$, 
where  
$H_0$ is the kinetic energy term,  
$H_0 = -t \sum
 c_{i\alpha}^\dagger  c_{j\alpha}$, 
with 
$t$ being the nearest-neighbor hopping and $c_{i\alpha}$,  $c_{i\alpha}^\dagger$  the destruction and creation operators for an electron with spin projection $\alpha$ in site $i$ and $j$, respectively. The Rashba SOI contribution is given by
\begin{equation} 
H_R= 
\frac{i \lambda_R}{a_{ cc}}\sum_{\substack {<i,j>\\\alpha,\beta}} c_{i\alpha}^\dagger  [(\bm{\sigma}  \times \bm{d}_{ij}) \cdot \bm{e_p}]_{\alpha \beta} c_{j\beta} \,\,,
\label{HR} 
\end{equation}
with  $\lambda_R$ being the Rashba SOI strength that can be tuned by the electric field intensity,  
$\bm{\sigma}$ are the Pauli spin matrices, $\bm{d}_{ij}$ the position vector between sites ${i}$ and ${j}$, $\alpha, \beta$ are the spin projection indices. 
Due to the analytical expression for $H_R$ (Eq. \ref{HR}), a new symmetry occurs. If  $\lambda_R$ changes sign, the spin polarizations on the longitudinal and transversal directions also change sign, whereas the polarization in the perpendicular direction is unaffected, i.e., $P_{s}(-\lambda)= -P_{s}(\lambda), (s=t,l);  P_{s}(-\lambda)= P_{s}(\lambda), \, (s=p)$. Notice that the sign of $\lambda _R$ is determined by the sense of the electric field which originates it.

We consider  a graphene device  
composed of a central flake with Rashba SOI and two semiinfinite pristine nanoribbons of the same width as the conductor, see  Fig. \ref{device}. 
The conductance is computed in the Kubo approach by using the Green function formalism \cite{Xu2007,Diniz2012}. The spin-resolved  conductance  is given by
$\textit{G}^{LR}_{\sigma \sigma'} = \frac{e^2}{h} Tr[\Gamma^{L}_{\sigma}G^r_{\sigma,\sigma^\prime}\Gamma^{R}_{\sigma'}G^a_{\sigma',\sigma}]\,\,,$
where $G^{a(r)}_{\sigma, \sigma^\prime} $ is the advanced (retarded) Green function of the conductor and $\Gamma^{L(R)}_{\sigma}=i[\sum^{r}_{L(R),\sigma} - \sum^{a}_{L(R),\sigma}]$ is written in terms of the $L$ $(R)$ lead selfenergies $\Sigma^{a}_{L(R),\sigma}$.  In fact, an equivalent expression can be reached from a scattering approach, equivalent to the Kubo formalism, which evidences more clearly that the symmetry of the conductance matrix is the same as that of the Hamiltonian of the system. \cite{fisherlee,Chico1996b}

\subsection{Numerical results}

In order to verify the previous symmetry analysis, we perform numerical calculations for several graphene nanoribbons of similar widths and lengths, but for different symmetries and 
spin polarization directions. The most common expression for the Rashba Hamiltonian chooses $z$ as the electric field direction, and therefore sets this as
the spin quantization axis, even though it is not the most favorable from both the quantitative and the symmetry viewpoint. For some systems, such as  
TMC (WSe$_2$ and MoS$_2$), it has been experimentally reported that spin polarization can occur in this particular setup. Indeed, it has been dubbed a "Zeeman-type" spin splitting with an electric field, due to the fact that the spin and the external field are in the same direction.\cite{Yuan2013} 
We first analyze the results for 
this spin polarization direction, and then concentrate on the best configuration for the obtention of a spin-polarized current, namely, with the spin transversal to the
current and the electric field.

\subsubsection{Spin direction perpendicular to the plane of the ribbon.~~}

\begin{figure}
\includegraphics[width=0.9\columnwidth]{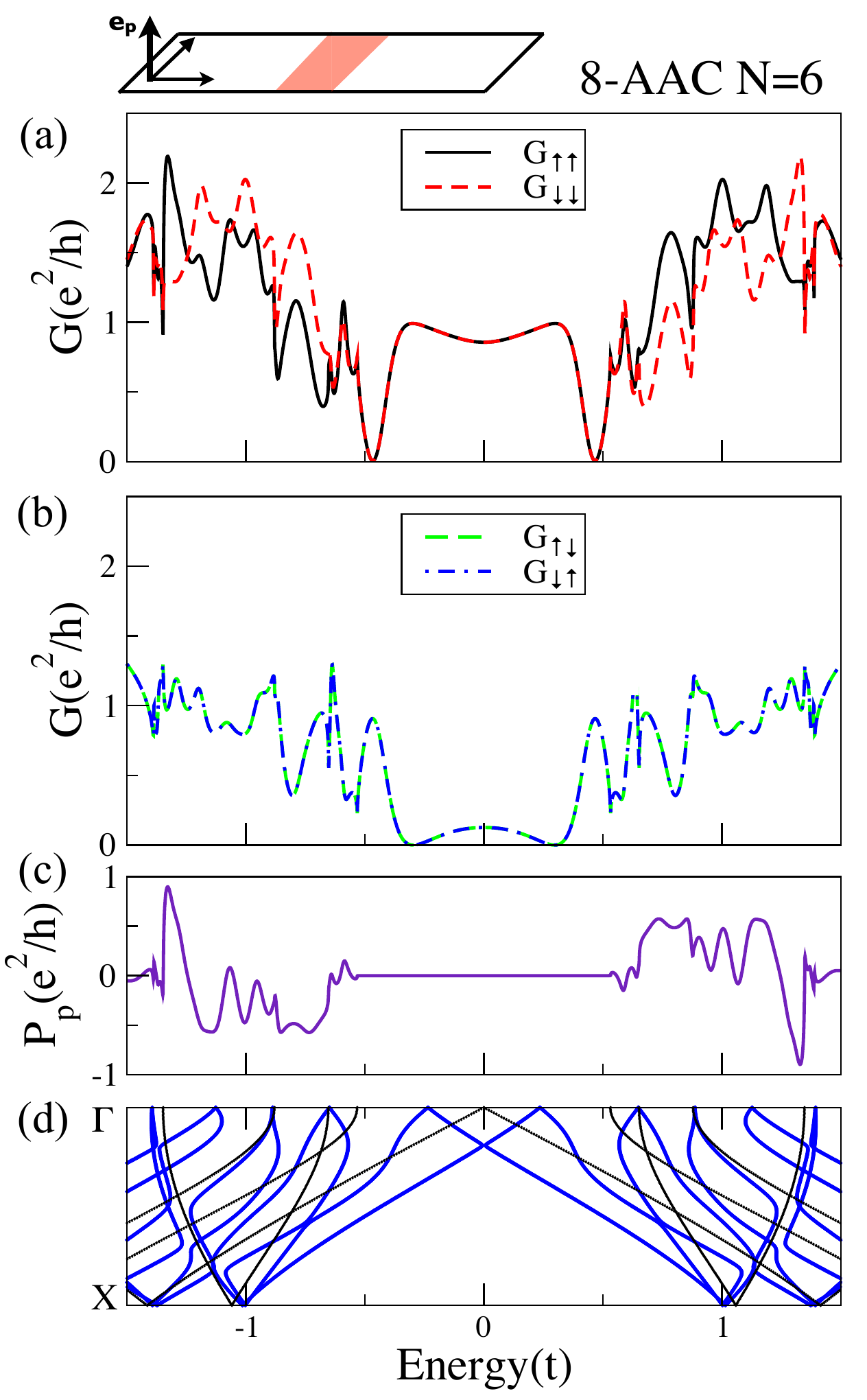}
\caption{Spin-resolved conductances and spin polarization of the current as a function of the energy for an 8-AAC GNR of length $N=6$  for spin direction perpendicular to the plane of the GNR and $\lambda_R=0.3t$. 
Panel (a) shows the spin-conserved and panel (b) the spin-flip conductances. Panel (c): Resulting spin polarization of the current as a function of the energy. 
In (d) we present the corresponding band structure of the infinite ribbon with Rashba SOI (thick blue lines) and without SOI (thin black lines). 
The top left schematic drawing indicates the spin projection direction used for this plot, shown with a thicker arrow. }
\label{figcondacp}
\end{figure}

When the spin direction is along $\bm{e_p}$, flakes AC and ZZ which have  $M_l$ symmetry cannot present a net spin-polarized current. 
We have checked numerically  this result, not shown here.
On the contrary,  the anti-armchair and anti-zigzag 
GNRs 
do not show $M_l$ symmetry,  
so they may have a net spin-polarized current. Indeed, we have found in both cases a finite spin-polarized transport; however, in this geometry the effect is small, especially for the anti-zigzag flakes. 
Therefore, for the sake of clarity of the spin-dependent conductance plots, we use a large value of SOI for this configuration, $\lambda_R=0.3t$. 
Fig. \ref{figcondacp} shows the spin-resolved conductances for an AAC graphene nanoribbon with $M=8$ and length $N=6$. 
The AAC flakes have $M_t$ symmetry and therefore  $\textit{G}^{LR}_{\uparrow\downarrow} = \textit{G}^{LR}_{\downarrow\uparrow}$, so a net  
spin-polarized current occurs because the spin-conserved conductances are different, 
$\textit{G}^{LR}_{\uparrow\uparrow} \neq \textit{G}^{LR}_{\downarrow\downarrow}$.

\begin{figure}
\includegraphics[width=0.9\columnwidth]{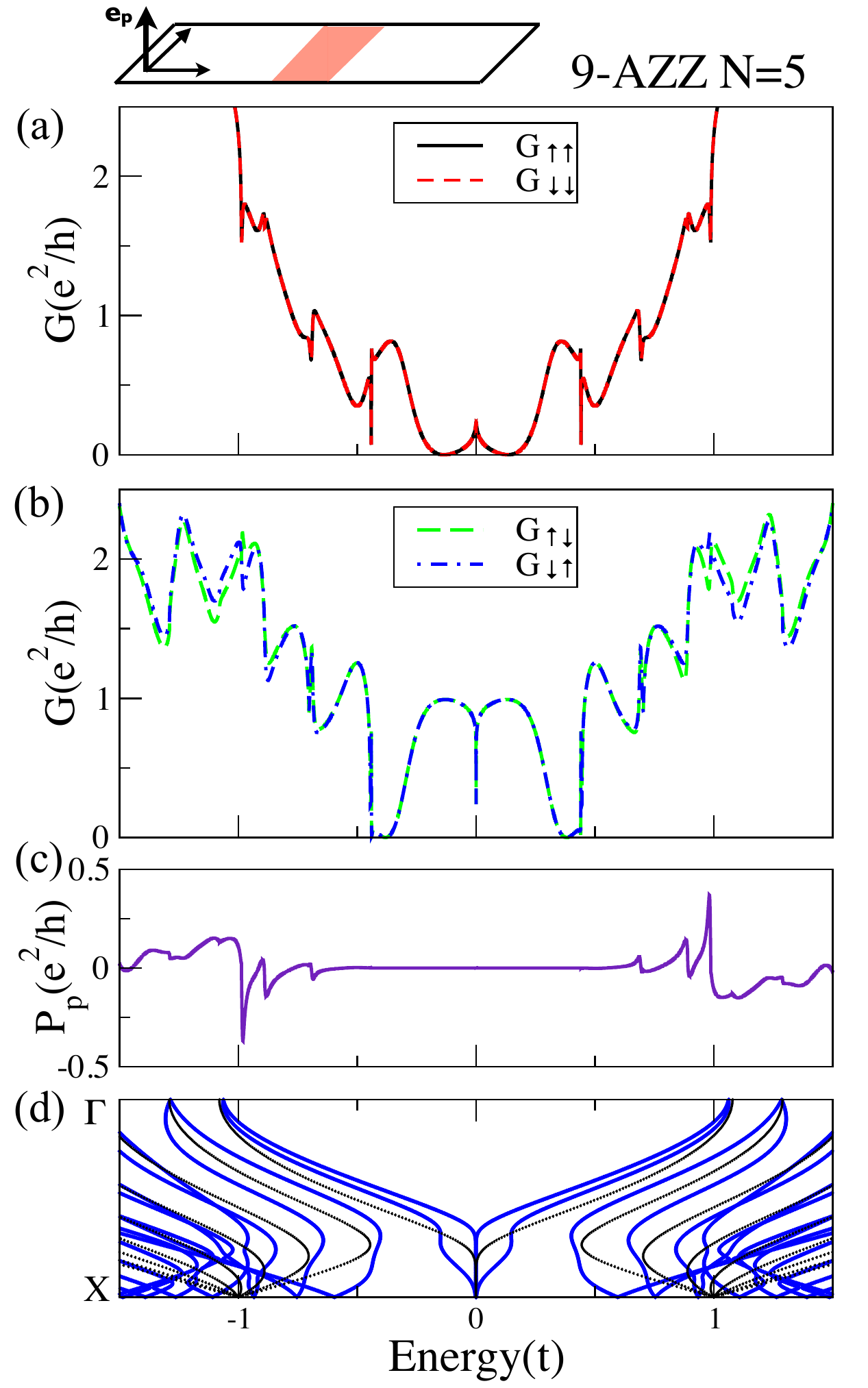}
\caption{Spin-resolved conductances and spin polarization of the current as a function of the energy for a 9-AZZ GNR with length $N=5$ and spin direction perpendicular to the plane of the GNR and $\lambda_R=0.3t$. (a) Spin-conserved conductances, which are equal due to $C_2$ symmetry; (b) spin-flip conductances, which give rise to the spin-polarized current. (c) Spin polarization of the current. (d) Band structure of the corresponding infinite ribbon calculated with Rashba SOI (thick blue lines) and without SO interaction.  The top left schematic drawing indicates the spin projection direction considered for this plot, shown with a thicker arrow. 
}
\label{figcondzzp}
\end{figure}

In contrast, as the AZZ case has $C_2$ symmetry, the spin-conserved conductances are equal and the spin polarization occurs because
$G^{LR}_{\uparrow\downarrow} \neq G^{LR}_{\downarrow\uparrow}$. Fig. \ref{figcondzzp} presents the spin-resolved conductances for an AZZ ribbon 
of width $M=9$ and length $N=5$. The spin current is smaller in this case. We attribute this to the fact that the infinite AZZ ribbon has also $M_t$ symmetry, absent in
the flake, which would yield equal spin-flip conductances. Therefore, in this case the spin-polarized current is clearly a finite-size or scattering effect, due to the boundary between 
the leads without Rashba and the flake with SOI.

In these two instances we can observe another relation for the conductances and polarization derived from both time reversal and electron-hole symmetries. It can be easily verified that the combination of $\Theta$ and $U$ (see Table \ref{TS}) yields $G^{LR}_{\sigma \sigma'} (E) = 
G^{LR}_{\bar{\sigma}\bar{\sigma}'}(-E)$. In terms of the polarization, this means $P_s(E) = -P_s(-E)$.  Figs. \ref{figcondacp} (a) and \ref{figcondzzp} (b) are non-trivial examples for the spin-conserved and spin-flip conductances, whereas Fig. \ref{figcondacp} (c) shows this symmetry in terms of the current polarization $P_p(E)$.

Notice that there is a region around $E_F$ where the spin polarization is zero. It corresponds to the energy range with only two bands in the energy spectrum, as it can be seen in panels (d) of Figs. \ref{figcondacp} and \ref{figcondzzp}. In order to have a net spin current, more than two bands should be available in the system.\cite{Liu2012}  Using wider ribbons lowers the energies of these additional bands.\cite{Zhang2013}

 \subsubsection{Spin projection parallel to the transversal direction of the nanoribbon.~~}

Now we discuss the case with {\it spin projection parallel to the nanoribbon transversal direction} $\bm e_t$. In this setup the spin-polarized conductance is the largest, so this configuration is the most relevant from the experimental viewpoint. This result can be inferred from the structure of the Rashba Hamiltonian. 
In the continuum model the Rashba term takes the form $H_R \propto ({\boldsymbol \sigma} 
\times {\bm k}) \cdot {\bm E}$, being $E$ the applied electric field.  
Thus, a maximum contribution can be expected when the directions of the electric field ${\bm E}$, the current direction ${\bm k}$ and the spin are orthogonal, as it happens when the spin is pointing in the transversal direction.
For this spin direction,  the symmetries do not impose any condition to the spin-conserved conductances in all considered cases, i.e., $\textit{G}^{LR}_{\uparrow\uparrow} \neq \textit{G}^{LR}_{\downarrow\downarrow}$. Actually, polarized currents are obtained in all four GNRs studied. 

\begin{figure*}[t]
\centering
\includegraphics[width=12cm]{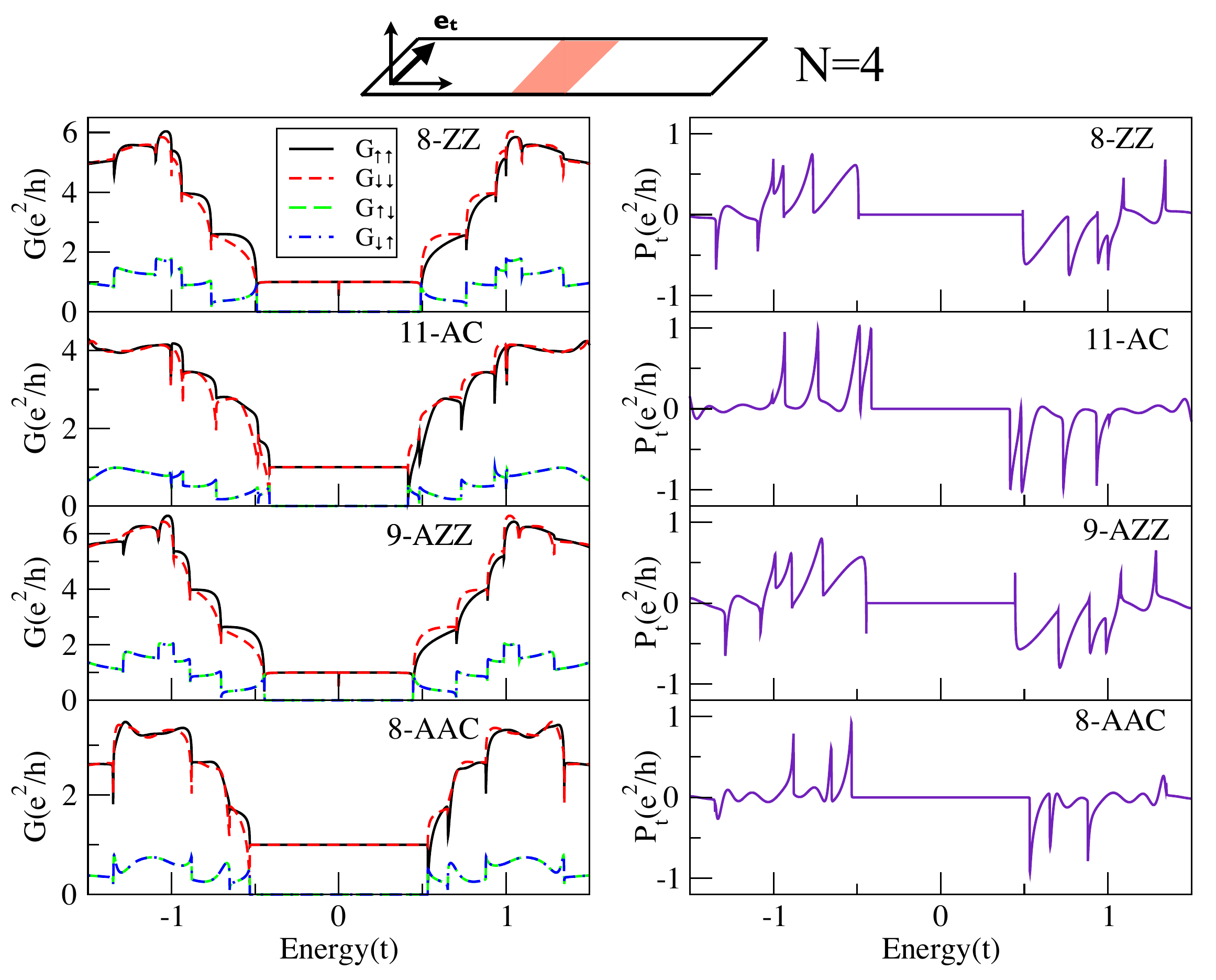}
\caption{ 
Left panels: Spin-resolved conductances as a function of the energy for an 8-ZZ, an 11-AC, a 9-AZZ and an 8-AAC graphene nanoribbons of length $N=4$, enumerated from top to bottom. In all cases the spin is projected in the transversal direction of the GNR, indicated with a thicker arrow in the schematic drawing at the top. 
Right panels show the corresponding polarization currents $P_t$. 
The results are obtained with $\lambda_R=0.1t$.}
\label{condsim}
\end{figure*}

Results for each one of the symmetries illustrated in Fig. \ref{figflakes}, namely, zigzag ($M=8$), armchair ($M=11$), anti-zigzag ($M=9$) and anti-armchair ($M=8$) GNRs are presented in Fig. \ref{condsim}.  The chosen 
ribbons have similar widths and the same fixed length, $N=4$. 
For this spin direction, $\bm{e_t}$, the magnitude of the polarized current is similar in the four cases. Moreover, the spin polarization of the symmetric GNRs presents magnitudes similar to those found in the asymmetric cases.  This result may be understood on the basis that $M_l$ does not play any role for this particular spin orientation. 
Also, it is evident the relation $P_s (E) = -P_s (-E)$, imposed by electron hole-symmetry, as given in Table \ref{TS}.

\subsubsection{Final remarks.~~}  

 Finally, we would like to mention that if the spin projection is taken along the longitudinal direction, the effect is small, although not zero. The particular conductance relations are also collected in the graphical summary presented in Fig. \ref{conductable}. 

As our main focus in this work was the role of symmetry in the spin-resolved conductances, we have chosen small flakes, for which the effects are clearer.
For longer flakes, the most notable (and obvious) difference would be that the number of conductance and polarization oscillations increases, due to the appearance of more quasi-localized states.

 It is interesting to mention that the set of symmetries discussed here also plays an important role on the size dependence of the spin polarization. As discussed previously, in some systems, the spin polarized currents arises because of a finite-size effect, whereas in others it is not so. If the polarization is due to a finite size effect, then we expect that its maximum value will eventually decrease with size, but in any case it should not grow on average. On the other hand, if the polarization is due to a lack of symmetry present both in the infinite and in the finite case, there should be a non-zero polarization for all sizes. 
 
 \begin{figure}
\includegraphics[width=\columnwidth]{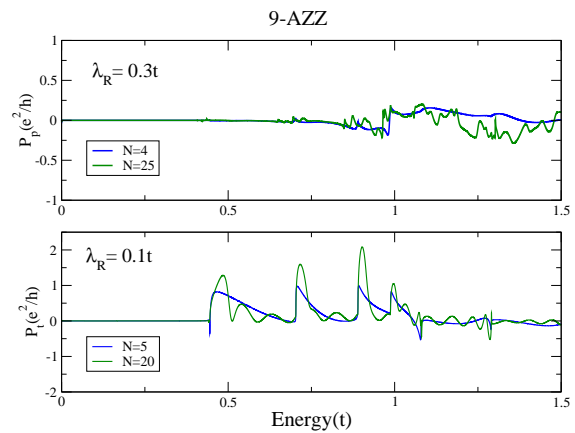}
\caption{Spin polarization currents as a function of the energy for  9-AZZ  GNRs with different Rashba SOI strength, spin polarization directions, and ribbon lengths: (a) N=5 and 20, $\lambda_R$=0.3$t$ and spin projected in the perpendicular direction ($P_p$) and (b) N=4 and 20,  $\lambda_R$=0.1$t$ and spin projected in the transversal direction ($P_t$).}
\label{figsize}
\end{figure}

As an example we have chosen a 9-AZZ ribbon; the corresponding flake has only $C_2$ symmetry, but the infinite ribbon has also $M_t$. For spin projection direction along $\bm{e_p}$, 
this implies an extra relation in the spin-resolved conductance (see Table \ref{TS}) that yields a zero polarization current. However, for spin projection direction along $\bm{e_t}$, the relation is the same as for $C_2$, so we expect the polarized current to exist for growing size. 
These size dependences are illustrated in Figure \ref{figsize}, where spin-polarized currents are shown for the two discussed spin polarization directions and two flake lengths with Rashba SOI.
  Besides the aforementioned oscillations due to the larger size, it is notable the increase of the polarization of the current presented in the bottom panel for the longer flake length.

As a general result, we have shown that for some particular graphene flakes (Fig. \ref{figflakes}), the use of symmetries allows us to elucidate which spin-resolved conductances are equal and which are different. In the same way, although not presented here, we can also infer that GNRs with symmetric chiral edges, as those obtained by opening carbon nanotubes, will behave as AZZ ribbons, because of their $C_2$ symmetry. With respect to the size dependence, for wider ribbons the number of bands increases, and the onset of spin polarized currents happens for lower energies because of the availability of more than two spin channels.

\section{Summary}

We have studied the symmetries of the spin-resolved conductances in planar devices with Rashba SOI. 
The combination of spatial mirror reflections and $C_2$ rotation with time-reversal symmetry leads to specific predictions with respect to the possibility of obtaining spin-polarized currents in such devices. 
As an example, we compute the spin-dependent transport of graphene nanoribbons with an applied electric field in finite region. We have shown that spin-polarized currents can be achieved if the spin polarization is measured in the transversal direction of the ribbon for all the ribbon geometries.  Furthermore, we have analyzed 
all the basic symmetries and spin directions, elucidating which configurations can yield a spin-polarized current on the basis of symmetry.  
The intensity and sign of the Rashba spin-orbit coupling may be modified by external electric field, opening the possibility of building an all-electrical spin valve.
Our findings can be useful for a smart design of spintronic graphene devices, being of general application to other materials with Rashba SOI.

\section*{Acknowledgements}

We  thank C. Lewenkopf, L. Lima, A. T. da Costa and M. Pelc for interesting discussions. This work has been financially supported by CNPq and FAPERJ under grant E-26/101.522/2010 and by Spanish MINECO grant No. FIS2012-33521. We acknowledge the financial support of the CNPq/CSIC project 2011BR0087 and of the INCT de Nanomateriais de carbono.


\begin{thebibliography}{46}
\makeatletter
\providecommand \@ifxundefined [1]{%
 \@ifx{#1\undefined}
}%
\providecommand \@ifnum [1]{%
 \ifnum #1\expandafter \@firstoftwo
 \else \expandafter \@secondoftwo
 \fi
}%
\providecommand \@ifx [1]{%
 \ifx #1\expandafter \@firstoftwo
 \else \expandafter \@secondoftwo
 \fi
}%
\providecommand \natexlab [1]{#1}%
\providecommand \enquote  [1]{``#1''}%
\providecommand \bibnamefont  [1]{#1}%
\providecommand \bibfnamefont [1]{#1}%
\providecommand \citenamefont [1]{#1}%
\providecommand \href@noop [0]{\@secondoftwo}%
\providecommand \href [0]{\begingroup \@sanitize@url \@href}%
\providecommand \@href[1]{\@@startlink{#1}\@@href}%
\providecommand \@@href[1]{\endgroup#1\@@endlink}%
\providecommand \@sanitize@url [0]{\catcode `\\12\catcode `\$12\catcode
  `\&12\catcode `\#12\catcode `\^12\catcode `\_12\catcode `\%12\relax}%
\providecommand \@@startlink[1]{}%
\providecommand \@@endlink[0]{}%
\providecommand \url  [0]{\begingroup\@sanitize@url \@url }%
\providecommand \@url [1]{\endgroup\@href {#1}{\urlprefix }}%
\providecommand \urlprefix  [0]{URL }%
\providecommand \Eprint [0]{\href }%
\providecommand \doibase [0]{http://dx.doi.org/}%
\providecommand \selectlanguage [0]{\@gobble}%
\providecommand \bibinfo  [0]{\@secondoftwo}%
\providecommand \bibfield  [0]{\@secondoftwo}%
\providecommand \translation [1]{[#1]}%
\providecommand \BibitemOpen [0]{}%
\providecommand \bibitemStop [0]{}%
\providecommand \bibitemNoStop [0]{.\EOS\space}%
\providecommand \EOS [0]{\spacefactor3000\relax}%
\providecommand \BibitemShut  [1]{\csname bibitem#1\endcsname}%
\let\auto@bib@innerbib\@empty
\bibitem [{\citenamefont {Wolf}\ \emph {et~al.}(2001)\citenamefont {Wolf},
  \citenamefont {Awschalom}, \citenamefont {Buhrman}, \citenamefont {Daughton},
  \citenamefont {von Moln\'ar}, \citenamefont {Roukes}, \citenamefont
  {Chtchelkanova},\ and\ \citenamefont {Treger}}]{Wolf2001}%
  \BibitemOpen
  \bibfield  {author} {\bibinfo {author} {\bibfnamefont {S.~A.}\ \bibnamefont
  {Wolf}}, \bibinfo {author} {\bibfnamefont {D.~D.}\ \bibnamefont {Awschalom}},
  \bibinfo {author} {\bibfnamefont {R.~A.}\ \bibnamefont {Buhrman}}, \bibinfo
  {author} {\bibfnamefont {J.~M.}\ \bibnamefont {Daughton}}, \bibinfo {author}
  {\bibfnamefont {S.}~\bibnamefont {von Moln\'ar}}, \bibinfo {author}
  {\bibfnamefont {M.~L.}\ \bibnamefont {Roukes}}, \bibinfo {author}
  {\bibfnamefont {A.~Y.}\ \bibnamefont {Chtchelkanova}}, \ and\ \bibinfo
  {author} {\bibfnamefont {D.~M.}\ \bibnamefont {Treger}},\ }\href {\doibase
  10.1126/science.1065389} {\bibfield  {journal} {\bibinfo  {journal}
  {Science}\ }\textbf {\bibinfo {volume} {294}},\ \bibinfo {pages} {1488}
  (\bibinfo {year} {2001})}\BibitemShut {NoStop}%
\bibitem [{\citenamefont {\ifmmode \check{Z}\else
  \v{Z}\fi{}uti\ifmmode~\acute{c}\else \'{c}\fi{}}\ \emph
  {et~al.}(2004)\citenamefont {\ifmmode \check{Z}\else
  \v{Z}\fi{}uti\ifmmode~\acute{c}\else \'{c}\fi{}}, \citenamefont {Fabian},\
  and\ \citenamefont {Das~Sarma}}]{Zutic2004}%
  \BibitemOpen
  \bibfield  {author} {\bibinfo {author} {\bibfnamefont {I.}~\bibnamefont
  {\ifmmode \check{Z}\else \v{Z}\fi{}uti\ifmmode~\acute{c}\else \'{c}\fi{}}},
  \bibinfo {author} {\bibfnamefont {J.}~\bibnamefont {Fabian}}, \ and\ \bibinfo
  {author} {\bibfnamefont {S.}~\bibnamefont {Das~Sarma}},\ }\href {\doibase
  10.1103/RevModPhys.76.323} {\bibfield  {journal} {\bibinfo  {journal} {Rev.
  Mod. Phys.}\ }\textbf {\bibinfo {volume} {76}},\ \bibinfo {pages} {323}
  (\bibinfo {year} {2004})}\BibitemShut {NoStop}%
\bibitem [{\citenamefont {Han}\ \emph {et~al.}(2014)\citenamefont {Han},
  \citenamefont {Kawakami}, \citenamefont {Gmitra},\ and\ \citenamefont
  {Fabian}}]{Han2014}%
  \BibitemOpen
  \bibfield  {author} {\bibinfo {author} {\bibfnamefont {W.}~\bibnamefont
  {Han}}, \bibinfo {author} {\bibfnamefont {R.~K.}\ \bibnamefont {Kawakami}},
  \bibinfo {author} {\bibfnamefont {M.}~\bibnamefont {Gmitra}}, \ and\ \bibinfo
  {author} {\bibfnamefont {J.}~\bibnamefont {Fabian}},\ }\href@noop {}
  {\bibfield  {journal} {\bibinfo  {journal} {Nat. Nanotech.}\ }\textbf
  {\bibinfo {volume} {9}},\ \bibinfo {pages} {794} (\bibinfo {year}
  {2014})}\BibitemShut {NoStop}%
\bibitem [{\citenamefont {Tang}\ and\ \citenamefont {Wang}(2015)}]{Tang2015}%
  \BibitemOpen
  \bibfield  {author} {\bibinfo {author} {\bibfnamefont {J.}~\bibnamefont
  {Tang}}\ and\ \bibinfo {author} {\bibfnamefont {K.~L.}\ \bibnamefont
  {Wang}},\ }\href {\doibase 10.1039/C4NR07611G} {\bibfield  {journal}
  {\bibinfo  {journal} {Nanoscale}\ }\textbf {\bibinfo {volume} {7}},\ \bibinfo
  {pages} {4325} (\bibinfo {year} {2015})}\BibitemShut {NoStop}%
\bibitem [{\citenamefont {Debray}\ \emph {et~al.}(2009)\citenamefont {Debray},
  \citenamefont {Rahman}, \citenamefont {Wan}, \citenamefont {Newrock},
  \citenamefont {Cahay}, \citenamefont {Ngo}, \citenamefont {Ulloa},
  \citenamefont {Herbert}, \citenamefont {Muhammad},\ and\ \citenamefont
  {Johnson}}]{Debray2009}%
  \BibitemOpen
  \bibfield  {author} {\bibinfo {author} {\bibfnamefont {P.}~\bibnamefont
  {Debray}}, \bibinfo {author} {\bibfnamefont {S.~M.~S.}\ \bibnamefont
  {Rahman}}, \bibinfo {author} {\bibfnamefont {J.}~\bibnamefont {Wan}},
  \bibinfo {author} {\bibfnamefont {R.~S.}\ \bibnamefont {Newrock}}, \bibinfo
  {author} {\bibfnamefont {M.}~\bibnamefont {Cahay}}, \bibinfo {author}
  {\bibfnamefont {A.~T.}\ \bibnamefont {Ngo}}, \bibinfo {author} {\bibfnamefont
  {S.~E.}\ \bibnamefont {Ulloa}}, \bibinfo {author} {\bibfnamefont {S.~T.}\
  \bibnamefont {Herbert}}, \bibinfo {author} {\bibfnamefont {M.}~\bibnamefont
  {Muhammad}}, \ and\ \bibinfo {author} {\bibfnamefont {M.}~\bibnamefont
  {Johnson}},\ }\href {http://dx.doi.org/10.1038/nnano.2009.240} {\bibfield
  {journal} {\bibinfo  {journal} {Nat. Nanotech.}\ }\textbf {\bibinfo {volume}
  {4}},\ \bibinfo {pages} {759} (\bibinfo {year} {2009})}\BibitemShut {NoStop}%
\bibitem [{\citenamefont {R.Winkler}(2003)}]{WinklerBook}%
  \BibitemOpen
  \bibfield  {author} {\bibinfo {author} {\bibnamefont {R.Winkler}},\
  }\href@noop {} {\emph {\bibinfo {title} {Spin-Orbit Coupling Effects in
  Two-Dimensional Electron and Hole Systems.}}}\ (\bibinfo  {publisher}
  {Springer-Verlag Berlin},\ \bibinfo {year} {2003})\BibitemShut {NoStop}%
\bibitem [{\citenamefont {Yuan}\ \emph {et~al.}(2013)\citenamefont {Yuan},
  \citenamefont {Bahramy}, \citenamefont {Morimoto}, \citenamefont {Wu},
  \citenamefont {Nomura}, \citenamefont {Yang}, \citenamefont {Shimotani},
  \citenamefont {Suzuki}, \citenamefont {Toh}, \citenamefont {Kloc},
  \citenamefont {Xu}, \citenamefont {Arita}, \citenamefont {Nagaosa},\ and\
  \citenamefont {Iwasa}}]{Yuan2013}%
  \BibitemOpen
  \bibfield  {author} {\bibinfo {author} {\bibfnamefont {H.}~\bibnamefont
  {Yuan}}, \bibinfo {author} {\bibfnamefont {M.~S.}\ \bibnamefont {Bahramy}},
  \bibinfo {author} {\bibfnamefont {K.}~\bibnamefont {Morimoto}}, \bibinfo
  {author} {\bibfnamefont {S.}~\bibnamefont {Wu}}, \bibinfo {author}
  {\bibfnamefont {K.}~\bibnamefont {Nomura}}, \bibinfo {author} {\bibfnamefont
  {B.-J.}\ \bibnamefont {Yang}}, \bibinfo {author} {\bibfnamefont
  {H.}~\bibnamefont {Shimotani}}, \bibinfo {author} {\bibfnamefont
  {R.}~\bibnamefont {Suzuki}}, \bibinfo {author} {\bibfnamefont
  {M.}~\bibnamefont {Toh}}, \bibinfo {author} {\bibfnamefont {C.}~\bibnamefont
  {Kloc}}, \bibinfo {author} {\bibfnamefont {X.}~\bibnamefont {Xu}}, \bibinfo
  {author} {\bibfnamefont {R.}~\bibnamefont {Arita}}, \bibinfo {author}
  {\bibfnamefont {N.}~\bibnamefont {Nagaosa}}, \ and\ \bibinfo {author}
  {\bibfnamefont {Y.}~\bibnamefont {Iwasa}},\ }\href
  {http://dx.doi.org/10.1038/nphys2691} {\bibfield  {journal} {\bibinfo
  {journal} {Nat. Phys.}\ }\textbf {\bibinfo {volume} {9}},\ \bibinfo {pages}
  {563} (\bibinfo {year} {2013})}\BibitemShut {NoStop}%
\bibitem [{\citenamefont {Yuan}\ \emph {et~al.}(2014)\citenamefont {Yuan},
  \citenamefont {Wang}, \citenamefont {Lian}, \citenamefont {Zhang},
  \citenamefont {Fang}, \citenamefont {Shen}, \citenamefont {Xu}, \citenamefont
  {Xu}, \citenamefont {Zhang}, \citenamefont {Hwang},\ and\ \citenamefont
  {Cui}}]{Yuan2014}%
  \BibitemOpen
  \bibfield  {author} {\bibinfo {author} {\bibfnamefont {H.}~\bibnamefont
  {Yuan}}, \bibinfo {author} {\bibfnamefont {X.}~\bibnamefont {Wang}}, \bibinfo
  {author} {\bibfnamefont {B.}~\bibnamefont {Lian}}, \bibinfo {author}
  {\bibfnamefont {H.}~\bibnamefont {Zhang}}, \bibinfo {author} {\bibfnamefont
  {X.}~\bibnamefont {Fang}}, \bibinfo {author} {\bibfnamefont {B.}~\bibnamefont
  {Shen}}, \bibinfo {author} {\bibfnamefont {G.}~\bibnamefont {Xu}}, \bibinfo
  {author} {\bibfnamefont {Y.}~\bibnamefont {Xu}}, \bibinfo {author}
  {\bibfnamefont {S.-C.}\ \bibnamefont {Zhang}}, \bibinfo {author}
  {\bibfnamefont {H.~Y.}\ \bibnamefont {Hwang}}, \ and\ \bibinfo {author}
  {\bibfnamefont {Y.}~\bibnamefont {Cui}},\ }\href
  {http://dx.doi.org/10.1038/nnano.2014.183} {\bibfield  {journal} {\bibinfo
  {journal} {Nat. Nanotech.}\ }\textbf {\bibinfo {volume} {9}},\ \bibinfo
  {pages} {851} (\bibinfo {year} {2014})}\BibitemShut {NoStop}%
\bibitem [{\citenamefont {Yin}\ \emph {et~al.}(2013)\citenamefont {Yin},
  \citenamefont {Yuan}, \citenamefont {Wang}, \citenamefont {Liu},
  \citenamefont {Zhang}, \citenamefont {Tang}, \citenamefont {Xu},
  \citenamefont {Chen}, \citenamefont {Shimotani}, \citenamefont {Iwasa},
  \citenamefont {Chen}, \citenamefont {Ge},\ and\ \citenamefont
  {Shen}}]{Yin2013}%
  \BibitemOpen
  \bibfield  {author} {\bibinfo {author} {\bibfnamefont {C.}~\bibnamefont
  {Yin}}, \bibinfo {author} {\bibfnamefont {H.}~\bibnamefont {Yuan}}, \bibinfo
  {author} {\bibfnamefont {X.}~\bibnamefont {Wang}}, \bibinfo {author}
  {\bibfnamefont {S.}~\bibnamefont {Liu}}, \bibinfo {author} {\bibfnamefont
  {S.}~\bibnamefont {Zhang}}, \bibinfo {author} {\bibfnamefont
  {N.}~\bibnamefont {Tang}}, \bibinfo {author} {\bibfnamefont {F.}~\bibnamefont
  {Xu}}, \bibinfo {author} {\bibfnamefont {Z.}~\bibnamefont {Chen}}, \bibinfo
  {author} {\bibfnamefont {H.}~\bibnamefont {Shimotani}}, \bibinfo {author}
  {\bibfnamefont {Y.}~\bibnamefont {Iwasa}}, \bibinfo {author} {\bibfnamefont
  {Y.}~\bibnamefont {Chen}}, \bibinfo {author} {\bibfnamefont {W.}~\bibnamefont
  {Ge}}, \ and\ \bibinfo {author} {\bibfnamefont {B.}~\bibnamefont {Shen}},\
  }\href {\doibase 10.1021/nl400153p} {\bibfield  {journal} {\bibinfo
  {journal} {Nano Letters}\ }\textbf {\bibinfo {volume} {13}},\ \bibinfo
  {pages} {2024} (\bibinfo {year} {2013})}\BibitemShut {NoStop}%
\bibitem [{\citenamefont {Tsai}\ \emph {et~al.}(2013)\citenamefont {Tsai},
  \citenamefont {Huang}, \citenamefont {Chang}, \citenamefont {Lin},
  \citenamefont {Jeng},\ and\ \citenamefont {Bansil}}]{Tsai2013}%
  \BibitemOpen
  \bibfield  {author} {\bibinfo {author} {\bibfnamefont {W.-F.}\ \bibnamefont
  {Tsai}}, \bibinfo {author} {\bibfnamefont {C.-Y.}\ \bibnamefont {Huang}},
  \bibinfo {author} {\bibfnamefont {T.-R.}\ \bibnamefont {Chang}}, \bibinfo
  {author} {\bibfnamefont {H.}~\bibnamefont {Lin}}, \bibinfo {author}
  {\bibfnamefont {H.-T.}\ \bibnamefont {Jeng}}, \ and\ \bibinfo {author}
  {\bibfnamefont {A.}~\bibnamefont {Bansil}},\ }\href
  {http://dx.doi.org/10.1038/ncomms2525} {\bibfield  {journal} {\bibinfo
  {journal} {Nat. Commun.}\ }\textbf {\bibinfo {volume} {4}},\ \bibinfo {pages}
  {1500} (\bibinfo {year} {2013})}\BibitemShut {NoStop}%
\bibitem [{\citenamefont {Ma}\ \emph {et~al.}(2014)\citenamefont {Ma},
  \citenamefont {Dai}, \citenamefont {Yin}, \citenamefont {Jing},\ and\
  \citenamefont {Huang}}]{Ma2014}%
  \BibitemOpen
  \bibfield  {author} {\bibinfo {author} {\bibfnamefont {Y.}~\bibnamefont
  {Ma}}, \bibinfo {author} {\bibfnamefont {Y.}~\bibnamefont {Dai}}, \bibinfo
  {author} {\bibfnamefont {N.}~\bibnamefont {Yin}}, \bibinfo {author}
  {\bibfnamefont {T.}~\bibnamefont {Jing}}, \ and\ \bibinfo {author}
  {\bibfnamefont {B.}~\bibnamefont {Huang}},\ }\href {\doibase
  10.1039/C4TC01394H} {\bibfield  {journal} {\bibinfo  {journal} {J. Mater.
  Chem. C}\ }\textbf {\bibinfo {volume} {2}},\ \bibinfo {pages} {8539}
  (\bibinfo {year} {2014})}\BibitemShut {NoStop}%
\bibitem [{\citenamefont {Liang}\ and\ \citenamefont {Gao}(2012)}]{Liang2012}%
  \BibitemOpen
  \bibfield  {author} {\bibinfo {author} {\bibfnamefont {D.}~\bibnamefont
  {Liang}}\ and\ \bibinfo {author} {\bibfnamefont {X.~P.}\ \bibnamefont
  {Gao}},\ }\href {\doibase 10.1021/nl301325h} {\bibfield  {journal} {\bibinfo
  {journal} {Nano Letters}\ }\textbf {\bibinfo {volume} {12}},\ \bibinfo
  {pages} {3263} (\bibinfo {year} {2012})}\BibitemShut {NoStop}%
\bibitem [{\citenamefont {Reuther}\ \emph {et~al.}(2013)\citenamefont
  {Reuther}, \citenamefont {Alicea},\ and\ \citenamefont
  {Yacoby}}]{Reuther2013}%
  \BibitemOpen
  \bibfield  {author} {\bibinfo {author} {\bibfnamefont {J.}~\bibnamefont
  {Reuther}}, \bibinfo {author} {\bibfnamefont {J.}~\bibnamefont {Alicea}}, \
  and\ \bibinfo {author} {\bibfnamefont {A.}~\bibnamefont {Yacoby}},\ }\href
  {\doibase 10.1103/PhysRevX.3.031011} {\bibfield  {journal} {\bibinfo
  {journal} {Phys. Rev. X}\ }\textbf {\bibinfo {volume} {3}},\ \bibinfo {pages}
  {031011} (\bibinfo {year} {2013})}\BibitemShut {NoStop}%
\bibitem [{\citenamefont {Zhai}\ and\ \citenamefont {Xu}(2005)}]{Feng2005}%
  \BibitemOpen
  \bibfield  {author} {\bibinfo {author} {\bibfnamefont {F.}~\bibnamefont
  {Zhai}}\ and\ \bibinfo {author} {\bibfnamefont {H.~Q.}\ \bibnamefont {Xu}},\
  }\href@noop {} {\bibfield  {journal} {\bibinfo  {journal} {Phys. Rev. Lett.}\
  }\textbf {\bibinfo {volume} {94}},\ \bibinfo {pages} {246601} (\bibinfo
  {year} {2005})}\BibitemShut {NoStop}%
\bibitem [{\citenamefont {Zhang}(2007)}]{Zhang2007}%
  \BibitemOpen
  \bibfield  {author} {\bibinfo {author} {\bibfnamefont {Z.-Y.}\ \bibnamefont
  {Zhang}},\ }\href {http://stacks.iop.org/0953-8984/19/i=1/a=016209}
  {\bibfield  {journal} {\bibinfo  {journal} {J. Phys. Condens. Matter}\
  }\textbf {\bibinfo {volume} {19}},\ \bibinfo {pages} {016209} (\bibinfo
  {year} {2007})}\BibitemShut {NoStop}%
\bibitem [{\citenamefont {Xu}\ \emph {et~al.}(2013)\citenamefont {Xu},
  \citenamefont {Xiao},\ and\ \citenamefont {Chen}}]{Xu2013}%
  \BibitemOpen
  \bibfield  {author} {\bibinfo {author} {\bibfnamefont {Z.}~\bibnamefont
  {Xu}}, \bibinfo {author} {\bibfnamefont {X.}~\bibnamefont {Xiao}}, \ and\
  \bibinfo {author} {\bibfnamefont {Y.}~\bibnamefont {Chen}},\ }\href {\doibase
  http://dx.doi.org/10.1016/j.physleta.2012.11.039} {\bibfield  {journal}
  {\bibinfo  {journal} {Phys. Lett. A}\ }\textbf {\bibinfo {volume} {377}},\
  \bibinfo {pages} {412} (\bibinfo {year} {2013})}\BibitemShut {NoStop}%
\bibitem [{\citenamefont {Klinovaja}\ and\ \citenamefont
  {Loss}(2013)}]{Jelena2013}%
  \BibitemOpen
  \bibfield  {author} {\bibinfo {author} {\bibfnamefont {J.}~\bibnamefont
  {Klinovaja}}\ and\ \bibinfo {author} {\bibfnamefont {D.}~\bibnamefont
  {Loss}},\ }\href {\doibase 10.1103/PhysRevX.3.011008} {\bibfield  {journal}
  {\bibinfo  {journal} {Phys. Rev. X}\ }\textbf {\bibinfo {volume} {3}},\
  \bibinfo {pages} {011008} (\bibinfo {year} {2013})}\BibitemShut {NoStop}%
\bibitem [{\citenamefont {Alomar}\ and\ \citenamefont
  {S\'anchez}(2014)}]{David2014}%
  \BibitemOpen
  \bibfield  {author} {\bibinfo {author} {\bibfnamefont {M.~I.}\ \bibnamefont
  {Alomar}}\ and\ \bibinfo {author} {\bibfnamefont {D.}~\bibnamefont
  {S\'anchez}},\ }\href {\doibase 10.1103/PhysRevB.89.115422} {\bibfield
  {journal} {\bibinfo  {journal} {Phys. Rev. B}\ }\textbf {\bibinfo {volume}
  {89}},\ \bibinfo {pages} {115422} (\bibinfo {year} {2014})}\BibitemShut
  {NoStop}%
\bibitem [{\citenamefont {Liu}\ \emph {et~al.}(2012)\citenamefont {Liu},
  \citenamefont {Chan},\ and\ \citenamefont {Wang}}]{Liu2012}%
  \BibitemOpen
  \bibfield  {author} {\bibinfo {author} {\bibfnamefont {J.-F.}\ \bibnamefont
  {Liu}}, \bibinfo {author} {\bibfnamefont {K.~S.}\ \bibnamefont {Chan}}, \
  and\ \bibinfo {author} {\bibfnamefont {J.}~\bibnamefont {Wang}},\ }\href
  {http://stacks.iop.org/0957-4484/23/i=9/a=095201} {\bibfield  {journal}
  {\bibinfo  {journal} {Nanotechnology}\ }\textbf {\bibinfo {volume} {23}},\
  \bibinfo {pages} {095201} (\bibinfo {year} {2012})}\BibitemShut {NoStop}%
\bibitem [{\citenamefont {Zhang}\ \emph {et~al.}(2013)\citenamefont {Zhang},
  \citenamefont {Chan}, \citenamefont {Lin},\ and\ \citenamefont
  {Liu}}]{Zhang2013}%
  \BibitemOpen
  \bibfield  {author} {\bibinfo {author} {\bibfnamefont {Q.}~\bibnamefont
  {Zhang}}, \bibinfo {author} {\bibfnamefont {K.}~\bibnamefont {Chan}},
  \bibinfo {author} {\bibfnamefont {Z.}~\bibnamefont {Lin}}, \ and\ \bibinfo
  {author} {\bibfnamefont {J.-F.}\ \bibnamefont {Liu}},\ }\href {\doibase
  http://dx.doi.org/10.1016/j.physleta.2012.12.032} {\bibfield  {journal}
  {\bibinfo  {journal} {Phys. Lett. A}\ }\textbf {\bibinfo {volume} {377}},\
  \bibinfo {pages} {632 } (\bibinfo {year} {2013})}\BibitemShut {NoStop}%
\bibitem [{\citenamefont {Huang}\ \emph {et~al.}(2012)\citenamefont {Huang},
  \citenamefont {Wei}, \citenamefont {Sun}, \citenamefont {Wong}, \citenamefont
  {Feng}, \citenamefont {Neto},\ and\ \citenamefont {Wee}}]{Huang2012}%
  \BibitemOpen
  \bibfield  {author} {\bibinfo {author} {\bibfnamefont {H.}~\bibnamefont
  {Huang}}, \bibinfo {author} {\bibfnamefont {D.}~\bibnamefont {Wei}}, \bibinfo
  {author} {\bibfnamefont {J.}~\bibnamefont {Sun}}, \bibinfo {author}
  {\bibfnamefont {S.~L.}\ \bibnamefont {Wong}}, \bibinfo {author}
  {\bibfnamefont {Y.~P.}\ \bibnamefont {Feng}}, \bibinfo {author}
  {\bibfnamefont {A.~H.~C.}\ \bibnamefont {Neto}}, \ and\ \bibinfo {author}
  {\bibfnamefont {A.~T.~S.}\ \bibnamefont {Wee}},\ }\href
  {http://dx.doi.org/10.1038/srep00983} {\bibfield  {journal} {\bibinfo
  {journal} {Sci. Rep.}\ }\textbf {\bibinfo {volume} {2}},\ \bibinfo {pages}
  {983} (\bibinfo {year} {2012})}\BibitemShut {NoStop}%
\bibitem [{\citenamefont {Baringhaus}\ \emph {et~al.}(2014)\citenamefont
  {Baringhaus}, \citenamefont {Ruan}, \citenamefont {Edler}, \citenamefont
  {Tejeda}, \citenamefont {Sicot}, \citenamefont {Taleb-Ibrahimi},
  \citenamefont {Li}, \citenamefont {Jiang}, \citenamefont {Conrad},
  \citenamefont {Berger}, \citenamefont {Tegenkamp},\ and\ \citenamefont
  {de~Heer}}]{Baringhaus2014}%
  \BibitemOpen
  \bibfield  {author} {\bibinfo {author} {\bibfnamefont {J.}~\bibnamefont
  {Baringhaus}}, \bibinfo {author} {\bibfnamefont {M.}~\bibnamefont {Ruan}},
  \bibinfo {author} {\bibfnamefont {F.}~\bibnamefont {Edler}}, \bibinfo
  {author} {\bibfnamefont {A.}~\bibnamefont {Tejeda}}, \bibinfo {author}
  {\bibfnamefont {M.}~\bibnamefont {Sicot}}, \bibinfo {author} {\bibfnamefont
  {A.}~\bibnamefont {Taleb-Ibrahimi}}, \bibinfo {author} {\bibfnamefont
  {A.-P.}\ \bibnamefont {Li}}, \bibinfo {author} {\bibfnamefont
  {Z.}~\bibnamefont {Jiang}}, \bibinfo {author} {\bibfnamefont {E.~H.}\
  \bibnamefont {Conrad}}, \bibinfo {author} {\bibfnamefont {C.}~\bibnamefont
  {Berger}}, \bibinfo {author} {\bibfnamefont {C.}~\bibnamefont {Tegenkamp}}, \
  and\ \bibinfo {author} {\bibfnamefont {W.~A.}\ \bibnamefont {de~Heer}},\
  }\href {\doibase 10.1038/nature12952} {\bibfield  {journal} {\bibinfo
  {journal} {Nature}\ }\textbf {\bibinfo {volume} {506}},\ \bibinfo {pages}
  {349} (\bibinfo {year} {2014})}\BibitemShut {NoStop}%
\bibitem [{\citenamefont {Huertas-Hernando}\ \emph {et~al.}(2006)\citenamefont
  {Huertas-Hernando}, \citenamefont {Guinea},\ and\ \citenamefont
  {Brataas}}]{Huertas2007}%
  \BibitemOpen
  \bibfield  {author} {\bibinfo {author} {\bibfnamefont {D.}~\bibnamefont
  {Huertas-Hernando}}, \bibinfo {author} {\bibfnamefont {F.}~\bibnamefont
  {Guinea}}, \ and\ \bibinfo {author} {\bibfnamefont {A.}~\bibnamefont
  {Brataas}},\ }\href@noop {} {\bibfield  {journal} {\bibinfo  {journal} {Phys.
  Rev. B}\ }\textbf {\bibinfo {volume} {74}},\ \bibinfo {pages} {155426}
  (\bibinfo {year} {2006})}\BibitemShut {NoStop}%
\bibitem [{\citenamefont {Min}\ \emph {et~al.}(2006)\citenamefont {Min},
  \citenamefont {Hill}, \citenamefont {Sinitsyn}, \citenamefont {Sahu},
  \citenamefont {Kleinman},\ and\ \citenamefont {MacDonald}}]{Min2007}%
  \BibitemOpen
  \bibfield  {author} {\bibinfo {author} {\bibfnamefont {H.}~\bibnamefont
  {Min}}, \bibinfo {author} {\bibfnamefont {J.}~\bibnamefont {Hill}}, \bibinfo
  {author} {\bibfnamefont {N.}~\bibnamefont {Sinitsyn}}, \bibinfo {author}
  {\bibfnamefont {B.}~\bibnamefont {Sahu}}, \bibinfo {author} {\bibfnamefont
  {L.}~\bibnamefont {Kleinman}}, \ and\ \bibinfo {author} {\bibfnamefont
  {A.}~\bibnamefont {MacDonald}},\ }\href {\doibase 10.1103/PhysRevB.74.165310}
  {\bibfield  {journal} {\bibinfo  {journal} {Phys. Rev. B}\ }\textbf {\bibinfo
  {volume} {74}},\ \bibinfo {pages} {165310} (\bibinfo {year}
  {2006})}\BibitemShut {NoStop}%
\bibitem [{\citenamefont {Castro~Neto}\ and\ \citenamefont
  {Guinea}(2009)}]{CastroNeto2009}%
  \BibitemOpen
  \bibfield  {author} {\bibinfo {author} {\bibfnamefont {A.}~\bibnamefont
  {Castro~Neto}}\ and\ \bibinfo {author} {\bibfnamefont {F.}~\bibnamefont
  {Guinea}},\ }\href {\doibase 10.1103/PhysRevLett.103.026804} {\bibfield
  {journal} {\bibinfo  {journal} {Phys. Rev. Lett.}\ }\textbf {\bibinfo
  {volume} {103}},\ \bibinfo {pages} {026804} (\bibinfo {year}
  {2009})}\BibitemShut {NoStop}%
\bibitem [{\citenamefont {Weeks}\ \emph {et~al.}(2011)\citenamefont {Weeks},
  \citenamefont {Hu}, \citenamefont {Alicea}, \citenamefont {Franz},\ and\
  \citenamefont {Wu}}]{Weeks2011}%
  \BibitemOpen
  \bibfield  {author} {\bibinfo {author} {\bibfnamefont {C.}~\bibnamefont
  {Weeks}}, \bibinfo {author} {\bibfnamefont {J.}~\bibnamefont {Hu}}, \bibinfo
  {author} {\bibfnamefont {J.}~\bibnamefont {Alicea}}, \bibinfo {author}
  {\bibfnamefont {M.}~\bibnamefont {Franz}}, \ and\ \bibinfo {author}
  {\bibfnamefont {R.}~\bibnamefont {Wu}},\ }\href {\doibase
  10.1103/PhysRevX.1.021001} {\bibfield  {journal} {\bibinfo  {journal} {Phys.
  Rev. X}\ }\textbf {\bibinfo {volume} {1}},\ \bibinfo {pages} {021001}
  (\bibinfo {year} {2011})}\BibitemShut {NoStop}%
\bibitem [{\citenamefont {Gmitra}\ \emph {et~al.}(2013)\citenamefont {Gmitra},
  \citenamefont {Kochan},\ and\ \citenamefont {Fabian}}]{Gmitra2013}%
  \BibitemOpen
  \bibfield  {author} {\bibinfo {author} {\bibfnamefont {M.}~\bibnamefont
  {Gmitra}}, \bibinfo {author} {\bibfnamefont {D.}~\bibnamefont {Kochan}}, \
  and\ \bibinfo {author} {\bibfnamefont {J.}~\bibnamefont {Fabian}},\ }\href
  {\doibase 10.1103/PhysRevLett.110.246602} {\bibfield  {journal} {\bibinfo
  {journal} {Phys. Rev. Lett.}\ }\textbf {\bibinfo {volume} {110}},\ \bibinfo
  {pages} {246602} (\bibinfo {year} {2013})}\BibitemShut {NoStop}%
\bibitem [{\citenamefont {Eremeev}\ \emph {et~al.}(2014)\citenamefont
  {Eremeev}, \citenamefont {Nechaev}, \citenamefont {Echenique},\ and\
  \citenamefont {Chulkov}}]{Eremeev2014}%
  \BibitemOpen
  \bibfield  {author} {\bibinfo {author} {\bibfnamefont {S.~V.}\ \bibnamefont
  {Eremeev}}, \bibinfo {author} {\bibfnamefont {I.~A.}\ \bibnamefont
  {Nechaev}}, \bibinfo {author} {\bibfnamefont {P.~M.}\ \bibnamefont
  {Echenique}}, \ and\ \bibinfo {author} {\bibfnamefont {E.~V.}\ \bibnamefont
  {Chulkov}},\ }\href {http://dx.doi.org/10.1038/srep06900} {\bibfield
  {journal} {\bibinfo  {journal} {Sci. Rep.}\ }\textbf {\bibinfo {volume}
  {4}},\ \bibinfo {pages} {6900} (\bibinfo {year} {2014})}\BibitemShut
  {NoStop}%
\bibitem [{\citenamefont {Ma}\ \emph {et~al.}(2012)\citenamefont {Ma},
  \citenamefont {Li},\ and\ \citenamefont {Yang}}]{Ma2012}%
  \BibitemOpen
  \bibfield  {author} {\bibinfo {author} {\bibfnamefont {D.}~\bibnamefont
  {Ma}}, \bibinfo {author} {\bibfnamefont {Z.}~\bibnamefont {Li}}, \ and\
  \bibinfo {author} {\bibfnamefont {Z.}~\bibnamefont {Yang}},\ }\href {\doibase
  http://dx.doi.org/10.1016/j.carbon.2011.08.055} {\bibfield  {journal}
  {\bibinfo  {journal} {Carbon}\ }\textbf {\bibinfo {volume} {50}},\ \bibinfo
  {pages} {297} (\bibinfo {year} {2012})}\BibitemShut {NoStop}%
\bibitem [{\citenamefont {Marchenko}\ \emph {et~al.}(2012)\citenamefont
  {Marchenko}, \citenamefont {Varykhalov}, \citenamefont {Scholz},
  \citenamefont {Bihlmayer}, \citenamefont {Rashba}, \citenamefont {Rybkin},
  \citenamefont {Shikin},\ and\ \citenamefont {Rader}}]{Marchenko2012}%
  \BibitemOpen
  \bibfield  {author} {\bibinfo {author} {\bibfnamefont {D.}~\bibnamefont
  {Marchenko}}, \bibinfo {author} {\bibfnamefont {A.}~\bibnamefont
  {Varykhalov}}, \bibinfo {author} {\bibfnamefont {M.~R.}\ \bibnamefont
  {Scholz}}, \bibinfo {author} {\bibfnamefont {G.}~\bibnamefont {Bihlmayer}},
  \bibinfo {author} {\bibfnamefont {E.~I.}\ \bibnamefont {Rashba}}, \bibinfo
  {author} {\bibfnamefont {A.}~\bibnamefont {Rybkin}}, \bibinfo {author}
  {\bibfnamefont {A.~M.}\ \bibnamefont {Shikin}}, \ and\ \bibinfo {author}
  {\bibfnamefont {O.}~\bibnamefont {Rader}},\ }\href
  {http://dx.doi.org/10.1038/ncomms2227} {\bibfield  {journal} {\bibinfo
  {journal} {Nat. Commun.}\ }\textbf {\bibinfo {volume} {3}},\ \bibinfo {pages}
  {1232} (\bibinfo {year} {2012})}\BibitemShut {NoStop}%
\bibitem [{\citenamefont {Balakrishnan}\ \emph {et~al.}(2013)\citenamefont
  {Balakrishnan}, \citenamefont {Kok Wai~Koon}, \citenamefont {Jaiswal},
  \citenamefont {Castro~Neto},\ and\ \citenamefont
  {Ozyilmaz}}]{Balakrishnan2013}%
  \BibitemOpen
  \bibfield  {author} {\bibinfo {author} {\bibfnamefont {J.}~\bibnamefont
  {Balakrishnan}}, \bibinfo {author} {\bibfnamefont {G.}~\bibnamefont {Kok
  Wai~Koon}}, \bibinfo {author} {\bibfnamefont {M.}~\bibnamefont {Jaiswal}},
  \bibinfo {author} {\bibfnamefont {A.~H.}\ \bibnamefont {Castro~Neto}}, \ and\
  \bibinfo {author} {\bibfnamefont {B.}~\bibnamefont {Ozyilmaz}},\ }\href
  {http://dx.doi.org/10.1038/nphys2576} {\bibfield  {journal} {\bibinfo
  {journal} {Nat. Phys.}\ }\textbf {\bibinfo {volume} {9}},\ \bibinfo {pages}
  {284} (\bibinfo {year} {2013})}\BibitemShut {NoStop}%
\bibitem [{\citenamefont {Avsar}\ \emph {et~al.}(2014)\citenamefont {Avsar},
  \citenamefont {Tan}, \citenamefont {Taychatanapat}, \citenamefont
  {Balakrishnan}, \citenamefont {Koon}, \citenamefont {Yeo}, \citenamefont
  {Lahiri}, \citenamefont {Carvalho}, \citenamefont {Rodin}, \citenamefont
  {O'Farrell}, \citenamefont {Eda}, \citenamefont {Castro~Neto},\ and\
  \citenamefont {{\"O}zyilmaz}}]{Avsar2014}%
  \BibitemOpen
  \bibfield  {author} {\bibinfo {author} {\bibfnamefont {A.}~\bibnamefont
  {Avsar}}, \bibinfo {author} {\bibfnamefont {J.~Y.}\ \bibnamefont {Tan}},
  \bibinfo {author} {\bibfnamefont {T.}~\bibnamefont {Taychatanapat}}, \bibinfo
  {author} {\bibfnamefont {J.}~\bibnamefont {Balakrishnan}}, \bibinfo {author}
  {\bibfnamefont {G.~K.~W.}\ \bibnamefont {Koon}}, \bibinfo {author}
  {\bibfnamefont {Y.}~\bibnamefont {Yeo}}, \bibinfo {author} {\bibfnamefont
  {J.}~\bibnamefont {Lahiri}}, \bibinfo {author} {\bibfnamefont
  {A.}~\bibnamefont {Carvalho}}, \bibinfo {author} {\bibfnamefont {A.~S.}\
  \bibnamefont {Rodin}}, \bibinfo {author} {\bibfnamefont {E.~C.~T.}\
  \bibnamefont {O'Farrell}}, \bibinfo {author} {\bibfnamefont {G.}~\bibnamefont
  {Eda}}, \bibinfo {author} {\bibfnamefont {A.~H.}\ \bibnamefont
  {Castro~Neto}}, \ and\ \bibinfo {author} {\bibfnamefont {B.}~\bibnamefont
  {{\"O}zyilmaz}},\ }\href {http://dx.doi.org/10.1038/ncomms5875} {\bibfield
  {journal} {\bibinfo  {journal} {Nat. Commun.}\ }\textbf {\bibinfo {volume}
  {5}},\ \bibinfo {pages} {4875} (\bibinfo {year} {2014})}\BibitemShut
  {NoStop}%
\bibitem [{\citenamefont {Calleja}\ \emph {et~al.}(2015)\citenamefont
  {Calleja}, \citenamefont {Ochoa}, \citenamefont {Garnica}, \citenamefont
  {Barja}, \citenamefont {Navarro}, \citenamefont {Black}, \citenamefont
  {Otrokov}, \citenamefont {Chulkov}, \citenamefont {Arnau}, \citenamefont
  {Vazquez~de Parga}, \citenamefont {Guinea},\ and\ \citenamefont
  {Miranda}}]{Calleja2015}%
  \BibitemOpen
  \bibfield  {author} {\bibinfo {author} {\bibfnamefont {F.}~\bibnamefont
  {Calleja}}, \bibinfo {author} {\bibfnamefont {H.}~\bibnamefont {Ochoa}},
  \bibinfo {author} {\bibfnamefont {M.}~\bibnamefont {Garnica}}, \bibinfo
  {author} {\bibfnamefont {S.}~\bibnamefont {Barja}}, \bibinfo {author}
  {\bibfnamefont {J.~J.}\ \bibnamefont {Navarro}}, \bibinfo {author}
  {\bibfnamefont {A.}~\bibnamefont {Black}}, \bibinfo {author} {\bibfnamefont
  {M.~M.}\ \bibnamefont {Otrokov}}, \bibinfo {author} {\bibfnamefont {E.~V.}\
  \bibnamefont {Chulkov}}, \bibinfo {author} {\bibfnamefont {A.}~\bibnamefont
  {Arnau}}, \bibinfo {author} {\bibfnamefont {A.~L.}\ \bibnamefont {Vazquez~de
  Parga}}, \bibinfo {author} {\bibfnamefont {F.}~\bibnamefont {Guinea}}, \ and\
  \bibinfo {author} {\bibfnamefont {R.}~\bibnamefont {Miranda}},\ }\href
  {http://dx.doi.org/10.1038/nphys3173} {\bibfield  {journal} {\bibinfo
  {journal} {Nat. Phys.}\ }\textbf {\bibinfo {volume} {11}},\ \bibinfo {pages}
  {43} (\bibinfo {year} {2015})}\BibitemShut {NoStop}%
\bibitem [{\citenamefont {Balakrishnan}\ \emph {et~al.}(2014)\citenamefont
  {Balakrishnan}, \citenamefont {Koon}, \citenamefont {Avsar}, \citenamefont
  {Ho}, \citenamefont {Lee}, \citenamefont {Jaiswal}, \citenamefont {Baeck},
  \citenamefont {Ahn}, \citenamefont {Ferreira}, \citenamefont {Cazalilla},
  \citenamefont {Neto},\ and\ \citenamefont {{\"O}zyilmaz}}]{Balakrishnan2014}%
  \BibitemOpen
  \bibfield  {author} {\bibinfo {author} {\bibfnamefont {J.}~\bibnamefont
  {Balakrishnan}}, \bibinfo {author} {\bibfnamefont {G.~K.~W.}\ \bibnamefont
  {Koon}}, \bibinfo {author} {\bibfnamefont {A.}~\bibnamefont {Avsar}},
  \bibinfo {author} {\bibfnamefont {Y.}~\bibnamefont {Ho}}, \bibinfo {author}
  {\bibfnamefont {J.~H.}\ \bibnamefont {Lee}}, \bibinfo {author} {\bibfnamefont
  {M.}~\bibnamefont {Jaiswal}}, \bibinfo {author} {\bibfnamefont {S.-J.}\
  \bibnamefont {Baeck}}, \bibinfo {author} {\bibfnamefont {J.-H.}\ \bibnamefont
  {Ahn}}, \bibinfo {author} {\bibfnamefont {A.}~\bibnamefont {Ferreira}},
  \bibinfo {author} {\bibfnamefont {M.~A.}\ \bibnamefont {Cazalilla}}, \bibinfo
  {author} {\bibfnamefont {A.~H.~C.}\ \bibnamefont {Neto}}, \ and\ \bibinfo
  {author} {\bibfnamefont {B.}~\bibnamefont {{\"O}zyilmaz}},\ }\href
  {http://dx.doi.org/10.1038/ncomms5748} {\bibfield  {journal} {\bibinfo
  {journal} {Nat. Commun.}\ }\textbf {\bibinfo {volume} {5}},\ \bibinfo {pages}
  {4748} (\bibinfo {year} {2014})}\BibitemShut {NoStop}%
\bibitem [{\citenamefont {Varykhalov}\ \emph {et~al.}(2008)\citenamefont
  {Varykhalov}, \citenamefont {S\'anchez-Barriga}, \citenamefont {Shikin},
  \citenamefont {Biswas}, \citenamefont {Vescovo}, \citenamefont {Rybkin},
  \citenamefont {Marchenko},\ and\ \citenamefont {Rader}}]{Varykhalov2008}%
  \BibitemOpen
  \bibfield  {author} {\bibinfo {author} {\bibfnamefont {A.}~\bibnamefont
  {Varykhalov}}, \bibinfo {author} {\bibfnamefont {J.}~\bibnamefont
  {S\'anchez-Barriga}}, \bibinfo {author} {\bibfnamefont {A.~M.}\ \bibnamefont
  {Shikin}}, \bibinfo {author} {\bibfnamefont {C.}~\bibnamefont {Biswas}},
  \bibinfo {author} {\bibfnamefont {E.}~\bibnamefont {Vescovo}}, \bibinfo
  {author} {\bibfnamefont {A.}~\bibnamefont {Rybkin}}, \bibinfo {author}
  {\bibfnamefont {D.}~\bibnamefont {Marchenko}}, \ and\ \bibinfo {author}
  {\bibfnamefont {O.}~\bibnamefont {Rader}},\ }\href {\doibase
  10.1103/PhysRevLett.101.157601} {\bibfield  {journal} {\bibinfo  {journal}
  {Phys. Rev. Lett.}\ }\textbf {\bibinfo {volume} {101}},\ \bibinfo {pages}
  {157601} (\bibinfo {year} {2008})}\BibitemShut {NoStop}%
\bibitem [{\citenamefont {Dedkov}\ \emph {et~al.}(2008)\citenamefont {Dedkov},
  \citenamefont {Fonin}, \citenamefont {R\"udiger},\ and\ \citenamefont
  {Laubschat}}]{Dedkov2008}%
  \BibitemOpen
  \bibfield  {author} {\bibinfo {author} {\bibfnamefont {Y.~S.}\ \bibnamefont
  {Dedkov}}, \bibinfo {author} {\bibfnamefont {M.}~\bibnamefont {Fonin}},
  \bibinfo {author} {\bibfnamefont {U.}~\bibnamefont {R\"udiger}}, \ and\
  \bibinfo {author} {\bibfnamefont {C.}~\bibnamefont {Laubschat}},\ }\href
  {\doibase 10.1103/PhysRevLett.100.107602} {\bibfield  {journal} {\bibinfo
  {journal} {Phys. Rev. Lett.}\ }\textbf {\bibinfo {volume} {100}},\ \bibinfo
  {pages} {107602} (\bibinfo {year} {2008})}\BibitemShut {NoStop}%
\bibitem [{\citenamefont {Chico}\ \emph {et~al.}(2004)\citenamefont {Chico},
  \citenamefont {L\'opez-Sancho},\ and\ \citenamefont {Mu\~{n}oz}}]{Chico2004}%
  \BibitemOpen
  \bibfield  {author} {\bibinfo {author} {\bibfnamefont {L.}~\bibnamefont
  {Chico}}, \bibinfo {author} {\bibfnamefont {M.~P.}\ \bibnamefont
  {L\'opez-Sancho}}, \ and\ \bibinfo {author} {\bibfnamefont {M.~C.}\
  \bibnamefont {Mu\~{n}oz}},\ }\href@noop {} {\bibfield  {journal} {\bibinfo
  {journal} {Phys. Rev. Lett.}\ }\textbf {\bibinfo {volume} {93}},\ \bibinfo
  {pages} {176402} (\bibinfo {year} {2004})}\BibitemShut {NoStop}%
\bibitem [{\citenamefont {Costa}\ \emph {et~al.}(2013)\citenamefont {Costa},
  \citenamefont {Ferreira}, \citenamefont {Hallam}, \citenamefont {Duesberg},\
  and\ \citenamefont {Castro~Neto}}]{Costa2013}%
  \BibitemOpen
  \bibfield  {author} {\bibinfo {author} {\bibfnamefont {A.~T.}\ \bibnamefont
  {Costa}}, \bibinfo {author} {\bibfnamefont {M.~S.}\ \bibnamefont {Ferreira}},
  \bibinfo {author} {\bibfnamefont {T.}~\bibnamefont {Hallam}}, \bibinfo
  {author} {\bibfnamefont {G.~S.}\ \bibnamefont {Duesberg}}, \ and\ \bibinfo
  {author} {\bibfnamefont {A.~H.}\ \bibnamefont {Castro~Neto}},\ }\href
  {http://stacks.iop.org/0295-5075/104/i=4/a=47001} {\bibfield  {journal}
  {\bibinfo  {journal} {EPL (Europhysics Letters)}\ }\textbf {\bibinfo {volume}
  {104}},\ \bibinfo {pages} {47001} (\bibinfo {year} {2013})}\BibitemShut
  {NoStop}%
\bibitem [{\citenamefont {Zhang}\ \emph {et~al.}(2014)\citenamefont {Zhang},
  \citenamefont {Li}, \citenamefont {Wu}, \citenamefont {Wang}, \citenamefont
  {Culcer}, \citenamefont {Kaxiras},\ and\ \citenamefont {Zhang}}]{Zhang2014}%
  \BibitemOpen
  \bibfield  {author} {\bibinfo {author} {\bibfnamefont {G.}~\bibnamefont
  {Zhang}}, \bibinfo {author} {\bibfnamefont {X.}~\bibnamefont {Li}}, \bibinfo
  {author} {\bibfnamefont {G.}~\bibnamefont {Wu}}, \bibinfo {author}
  {\bibfnamefont {J.}~\bibnamefont {Wang}}, \bibinfo {author} {\bibfnamefont
  {D.}~\bibnamefont {Culcer}}, \bibinfo {author} {\bibfnamefont
  {E.}~\bibnamefont {Kaxiras}}, \ and\ \bibinfo {author} {\bibfnamefont
  {Z.}~\bibnamefont {Zhang}},\ }\href {\doibase 10.1039/C3NR05284B} {\bibfield
  {journal} {\bibinfo  {journal} {Nanoscale}\ }\textbf {\bibinfo {volume}
  {6}},\ \bibinfo {pages} {3259} (\bibinfo {year} {2014})}\BibitemShut
  {NoStop}%
\bibitem [{\citenamefont {Brey}\ and\ \citenamefont {Fertig}(2006)}]{Brey2006}%
  \BibitemOpen
  \bibfield  {author} {\bibinfo {author} {\bibfnamefont {L.}~\bibnamefont
  {Brey}}\ and\ \bibinfo {author} {\bibfnamefont {H.~A.}\ \bibnamefont
  {Fertig}},\ }\href {\doibase 10.1103/PhysRevB.73.235411} {\bibfield
  {journal} {\bibinfo  {journal} {Phys. Rev. B}\ }\textbf {\bibinfo {volume}
  {73}},\ \bibinfo {pages} {235411} (\bibinfo {year} {2006})}\BibitemShut
  {NoStop}%
\bibitem [{\citenamefont {Qiao}\ \emph {et~al.}(2010)\citenamefont {Qiao},
  \citenamefont {Yang}, \citenamefont {Feng}, \citenamefont {Tse},
  \citenamefont {Ding}, \citenamefont {Yao}, \citenamefont {Wang},\ and\
  \citenamefont {Niu}}]{Qiao2010}%
  \BibitemOpen
  \bibfield  {author} {\bibinfo {author} {\bibfnamefont {Z.}~\bibnamefont
  {Qiao}}, \bibinfo {author} {\bibfnamefont {S.}~\bibnamefont {Yang}}, \bibinfo
  {author} {\bibfnamefont {W.}~\bibnamefont {Feng}}, \bibinfo {author}
  {\bibfnamefont {W.-K.}\ \bibnamefont {Tse}}, \bibinfo {author} {\bibfnamefont
  {J.}~\bibnamefont {Ding}}, \bibinfo {author} {\bibfnamefont {Y.}~\bibnamefont
  {Yao}}, \bibinfo {author} {\bibfnamefont {J.}~\bibnamefont {Wang}}, \ and\
  \bibinfo {author} {\bibfnamefont {Q.}~\bibnamefont {Niu}},\ }\href {\doibase
  10.1103/PhysRevB.82.161414} {\bibfield  {journal} {\bibinfo  {journal} {Phys.
  Rev. B}\ }\textbf {\bibinfo {volume} {82}},\ \bibinfo {pages} {161414}
  (\bibinfo {year} {2010})}\BibitemShut {NoStop}%
\bibitem [{\citenamefont {Lenz}\ \emph {et~al.}(2013)\citenamefont {Lenz},
  \citenamefont {Urban},\ and\ \citenamefont {Bercioux}}]{Lenz2013}%
  \BibitemOpen
  \bibfield  {author} {\bibinfo {author} {\bibfnamefont {L.}~\bibnamefont
  {Lenz}}, \bibinfo {author} {\bibfnamefont {D.~F.}\ \bibnamefont {Urban}}, \
  and\ \bibinfo {author} {\bibfnamefont {D.}~\bibnamefont {Bercioux}},\ }\href
  {http://dx.doi.org/10.1140/epjb/e2013-40760-4} {\bibfield  {journal}
  {\bibinfo  {journal} {Eur. Phys. J. B}\ }\textbf {\bibinfo {volume} {86}},\
  \bibinfo {pages} {502} (\bibinfo {year} {2013})}\BibitemShut {NoStop}%
\bibitem [{\citenamefont {Xu}\ \emph {et~al.}(2007)\citenamefont {Xu},
  \citenamefont {Li}, \citenamefont {Pan},\ and\ \citenamefont {Zhu}}]{Xu2007}%
  \BibitemOpen
  \bibfield  {author} {\bibinfo {author} {\bibfnamefont {F.}~\bibnamefont
  {Xu}}, \bibinfo {author} {\bibfnamefont {B.}~\bibnamefont {Li}}, \bibinfo
  {author} {\bibfnamefont {H.}~\bibnamefont {Pan}}, \ and\ \bibinfo {author}
  {\bibfnamefont {J.-L.}\ \bibnamefont {Zhu}},\ }\href {\doibase
  10.1103/PhysRevB.75.085431} {\bibfield  {journal} {\bibinfo  {journal} {Phys.
  Rev. B}\ }\textbf {\bibinfo {volume} {75}},\ \bibinfo {pages} {085431}
  (\bibinfo {year} {2007})}\BibitemShut {NoStop}%
\bibitem [{\citenamefont {Diniz}\ \emph {et~al.}(2012)\citenamefont {Diniz},
  \citenamefont {Latg\'e},\ and\ \citenamefont {Ulloa}}]{Diniz2012}%
  \BibitemOpen
  \bibfield  {author} {\bibinfo {author} {\bibfnamefont {G.~S.}\ \bibnamefont
  {Diniz}}, \bibinfo {author} {\bibfnamefont {A.}~\bibnamefont {Latg\'e}}, \
  and\ \bibinfo {author} {\bibfnamefont {S.~E.}\ \bibnamefont {Ulloa}},\ }\href
  {\doibase 10.1103/PhysRevLett.108.126601} {\bibfield  {journal} {\bibinfo
  {journal} {Phys. Rev. Lett.}\ }\textbf {\bibinfo {volume} {108}},\ \bibinfo
  {pages} {126601} (\bibinfo {year} {2012})}\BibitemShut {NoStop}%
\bibitem [{\citenamefont {Fisher}\ and\ \citenamefont {Lee}(1981)}]{fisherlee}%
  \BibitemOpen
  \bibfield  {author} {\bibinfo {author} {\bibfnamefont {D.}~\bibnamefont
  {Fisher}}\ and\ \bibinfo {author} {\bibfnamefont {P.}~\bibnamefont {Lee}},\
  }\href {\doibase 10.1103/PhysRevB.23.6851} {\bibfield  {journal} {\bibinfo
  {journal} {Phys. Rev. B}\ }\textbf {\bibinfo {volume} {23}},\ \bibinfo
  {pages} {6851} (\bibinfo {year} {1981})}\BibitemShut {NoStop}%
\bibitem [{\citenamefont {Chico}\ \emph {et~al.}(1996)\citenamefont {Chico},
  \citenamefont {Benedict}, \citenamefont {Louie},\ and\ \citenamefont
  {Cohen}}]{Chico1996b}%
  \BibitemOpen
  \bibfield  {author} {\bibinfo {author} {\bibfnamefont {L.}~\bibnamefont
  {Chico}}, \bibinfo {author} {\bibfnamefont {L.~X.}\ \bibnamefont {Benedict}},
  \bibinfo {author} {\bibfnamefont {S.~G.}\ \bibnamefont {Louie}}, \ and\
  \bibinfo {author} {\bibfnamefont {M.~L.}\ \bibnamefont {Cohen}},\ }\href@noop
  {} {\bibfield  {journal} {\bibinfo  {journal} {Phys. Rev. B}\ }\textbf
  {\bibinfo {volume} {54}},\ \bibinfo {pages} {2600} (\bibinfo {year}
  {1996})}\BibitemShut {NoStop}%
\end{thebibliography}

%

\end{document}